\documentclass[12pt]{article}
 \pdfoutput=1

\usepackage{amsmath,amssymb,amsfonts}
\usepackage{hyperref}
\usepackage{graphicx}

\newcommand{\be}{\begin{eqnarray}}
\newcommand{\ee}{\end{eqnarray}}
\newcommand{\cL}{\mathcal{L}}
\newcommand{\cO}{\mathcal{O}}
\newcommand {\unit} [1] {\textrm{ #1}}
\newcommand{\hp}{h^\prime}
\newcommand{\gd}{\gamma_d}
\newcommand{\br}{{\rm Br}}
\newcommand{\chit}{\tilde \chi}
\newcommand{\GoH}{\frac{\Gamma_{N_1}}{H_1}}
\newcommand{\HoG}{\frac{H_1}{\Gamma_{N_1}}}
\newcommand{\goh}{\Gamma_{N_1}/H_1}

\newcommand{\ba}{\begin{array}}  
\newcommand{\ea}{\end{array}}

\numberwithin{equation}{section}

\begin{document}
\begin{titlepage}
\vspace{-1cm}
\begin{flushright}
LPT-ORSAY 11-09
\end{flushright}
\vspace{0.2cm}
\begin{center}
{\huge \bf Asymmetric Dark Matter from Leptogenesis}
\vspace*{0.2cm}
\end{center}
\vskip0.2cm

\begin{center}
{\bf  Adam Falkowski$^{a}$, Joshua T.~Ruderman$^{b}$ and Tomer Volansky$^{c,d}$}

\end{center}
\vskip 8pt

\begin{center}
{\it $^{a}$ Laboratoire de Physique Th\'eorique d'Orsay, UMR8627--CNRS,\\
Universit\'e Paris--Sud, F--91405 Orsay Cedex, France}
\\
{\it $^{b}$ Department of Physics, Princeton University, 
Princeton, 
NJ 08544, USA}
\\
{\it $^{c}$ Berkeley Center for Theoretical Physics, Department of Physics,
\\University of California, Berkeley, CA 94720, USA}
\\
{\it $^{d}$Theoretical Physics Group, Lawrence Berkeley National Laboratory, 
\\Berkeley, CA 94720, USA}

\begin{center}
{\tt Emails: adam.falkowski@th.u-psud.fr, rudes@ias.edu, tomerv@post.tau.ac.il}
\end{center}

\vspace*{0.3cm}

\end{center}

\vglue 0.3truecm

\begin{abstract}
  \vskip 3pt \noindent

 We present a new realization of asymmetric dark matter in which the dark matter and lepton asymmetries are generated simultaneously through two-sector leptogenesis.
 The right-handed neutrinos couple both to the Standard Model and to a hidden sector where the dark matter resides.  This framework explains the lepton asymmetry, dark matter abundance and neutrino masses all at once.  In contrast to previous realizations of asymmetric dark matter, the model allows for a wide range of dark matter masses, from keV to 10 TeV\@.  In particular, very light dark matter can be accommodated without violating experimental constraints.    We discuss several variants of our model that highlight interesting  phenomenological possibilities.   In one, late decays repopulate the symmetric dark matter component, providing a new mechanism for generating a large annihilation rate at the present epoch and allowing for mixed warm/cold dark matter.   In a second scenario, dark matter mixes with the active neutrinos, thus presenting a distinct method to populate sterile neutrino dark matter through leptogenesis.   At late times, oscillations and dark matter decays lead to interesting indirect detection signals.

\end{abstract}

\end{titlepage}

\newpage
\tableofcontents

\section{Introduction}\label{sec:intro}

The neutrino masses, the baryon asymmetry of the universe, and the existence of Dark Matter (DM) are the three experimental facts that clearly point to physics beyond the Standard Model (SM)\@.  
Interestingly, it is plausible that all three are related. Ê
On the one hand, neutrino masses suggest the existence of heavy sterile neutrinos, whose decays in the early universe can naturally produce the baryon asymmetry via leptogenesis~\cite{Fukugita:1986hr}  (for reviews with further references see e.g.~\cite{Davidson:2008bu,Giudice:2003jh}). Ê
On the other hand, the baryon and DM energy densities are of the same order, $\Omega_{\rm DM}/\Omega_{\rm b} \sim 5$,  suggesting that they may have a common origin.

One framework that relates the baryon and DM relic densities is Asymmetric DM (ADM)~\cite{Nussinov:1985xr, Kaplan:1991ah}. 
In this framework, the DM particle is distinct from its antiparticle and carries a conserved quantum number.  
An asymmetry in the particle-antiparticle number densities  is generated in the early universe. 
Subsequently, the symmetric component is annihilated away by sufficiently fast CP-conserving interactions, leaving the asymmetric component to dominate the relic density.  
Thus, the relic DM abundance is determined by the asymmetry, rather than by the annihilation cross-section,  in close analogy to SM baryogenesis,  and in stark contrast  to the thermal DM scenario.

The ADM scenario has been extensively studied in the literature~\cite{Kaplan:2009ag,Cohen:2010kn,Kitano:2004sv,Shelton:2010ta}.  In many existing realizations, an asymmetry is first produced in  one sector, either in the SM or in the DM sector,  and is then transferred to the other sector at later times by contact interactions. 
Such a scenario typically predicts similar baryon and DM number densities, $n_{\rm DM} \sim n_{\rm b} $ (see however~\cite{Nussinov:1985xr,Cohen:2009fz,Buckley:2010ui}), which then allows one to explain the observed $\Omega_{\rm DM}/\Omega_{\rm b}$ ratio for dark matter mass in the GeV ballpark.  
However, the lepton and DM asymmetries may be produced simultaneously at a very high temperature. 
This scenario, which we refer  to as {\em two-sector leptogenesis}, may profoundly alter the standard ADM predictions.   
Several authors have  previously considered generating the DM asymmetry from leptogenesis  \cite{An:2009vq}, 
however in the context of more specific models  that predict $n_{\rm DM} \sim n_{\rm b}$.  
In this paper we define a general framework  for two-sector leptogenesis and demonstrate that it  may naturally lead to a large hierarchy between $n_{\rm DM}$ and $n_{\rm b}$.
Thus, a wide range of DM masses, from about 1 keV to $10$ TeV, can be obtained within the ADM paradigm.

\begin{figure}[t]
\begin{center}
\includegraphics[scale=0.75]{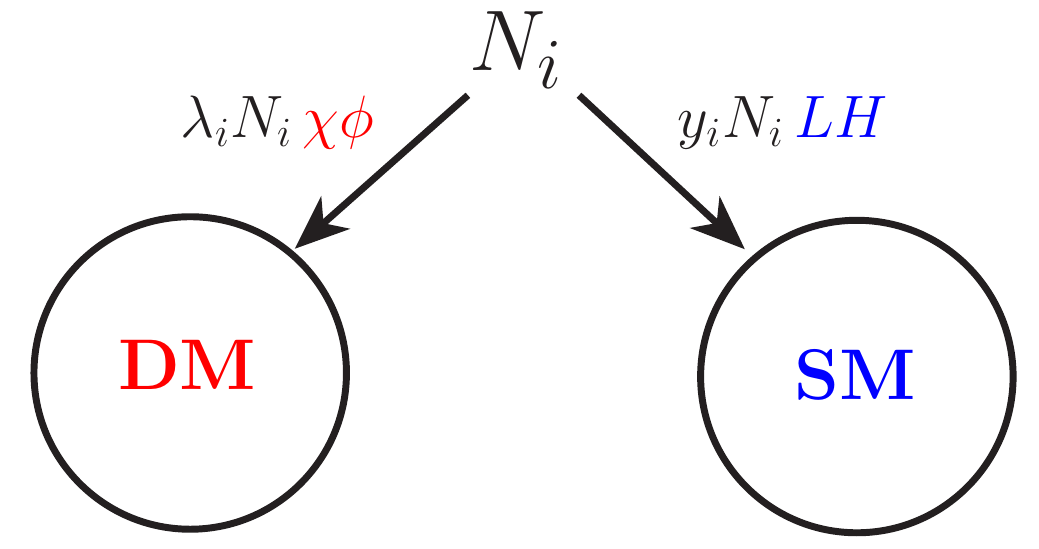}
\caption{\it\small A schematic view of our framework: the SM and DM sectors are indirectly connected via Yukawa interactions with the same heavy right-handed neutrinos, $N_i$.  The complex couplings, $\lambda_i$ and $y_i$, lead to CP violation in $N_i$ decays, and consequently particle-antiparticle asymmetries for DM and leptons.}
\label{fig:framework}
\end{center}
\end{figure}

The general framework we consider is sketched in Fig.~\ref{fig:framework}. Ê
DM resides in a hidden sector indirectly connected to the SM via Yukawa interactions with heavy Majorana neutrinos, $N_i$. 
In this set-up, the SM  leptons and the DM particle are charged under an approximate lepton number, which is broken by the Majorana masses of $N_i$. 
The Yukawa couplings can be  complex, leading to CP violation in the decays of $N_i$. 
Throughout this paper we work within the framework of thermal leptogenesis.  Resonant leptogenesis \cite{Pilaftsis:2003gt}, Dirac leptogenesis~\cite{Akhmedov:1998qx},  or soft leptogenesis~\cite{Grossman:2003jv}  could also be considered in this context and would be interesting to pursue.

The generation of the DM abundance adheres to the following steps, 
\begin{itemize}
\item A population of (at least) the lightest Majorana neutrino, $N_1$, is generated in the early universe.
\item At temperatures below $M_{N_1}$, these neutrinos decay out of equilibrium to both sectors.  Ê
The CP-violating decays lead to a lepton number asymmetry in both the SM and hidden sector.
\item As the universe cools well below $M_{N_1}$, the washout of lepton asymmetry, and its transfer between the 2 sectors, becomes inefficient and the asymmetries are frozen-in.
The asymptotic asymmetry can, in general, be different in the two sectors due to different branching fractions  and/or washout effects.  Ê
\item As usual, the SM lepton asymmetry is transferred into baryon asymmetry via electroweak sphalerons.  Ê
The symmetric baryon component is almost entirely wiped out by hadronic annihilations, and only the asymmetric component survives.  
\item  
Similarly, the symmetric component of the DM number density is annihilated away in the hidden sector. 
The relic abundance of DM is set by the remaining asymmetric component.  DM receives a Dirac mass,  $m_\chi \chi \chit$,  with another fermion state in the hidden sector, $\chit$.
\end{itemize}

We present a simple model that realizes the scenario described above. 
The hidden sector contains a chiral fermion $\chi$ - the DM candidate - and  a  complex scalar  $\phi$. 
The two are  coupled to heavy sterile neutrinos via Yukawa interactions,
$\lambda_i N_i \chi \phi$. 
We assume that any asymmetry carried by $\phi$ is immediately washed out by fast interactions, $\phi \leftrightarrow \phi^\dagger$, 
however only interactions mediated by the sterile neutrinos can turn $\chi$ into its antiparticle.
In this way, an asymmetry  $n_{\Delta \chi} = n_\chi - n_{\bar \chi}$ can survive at low temperatures below $M_{N_1}$. 
We also assume that additional interactions are present that allow the symmetric component of DM to annihilate.  For example, there may be a hidden $U(1)_d$ gauge symmetry, in which case $\chi$ and $\bar \chi$ can annihilate into hidden photons, $\chi + \bar \chi \rightarrow \gamma_d+ \gamma_d$, as in Ref.~\cite{Cohen:2010kn}.  
Within this framework we study the range of  asymmetries, in the SM and DM sectors, that can be generated during the 2-sector leptogenesis.  
In particular, we demonstrate that  the asymmetries in the two sectors may end up being vastly different,  
that is to say, $n_{\Delta \chi} \ll n_{\Delta L}$ or $n_{\Delta \chi} \gg n_{\Delta L}$.   
In such cases, the dark matter mass must be much larger, or much smaller, than a GeV, in order to recover $\Omega_{\rm DM} \sim \Omega_{b}$. 
We argue that, within this framework, DM masses in the  keV - 10 TeV range are easily obtained without violating any phenomenological constraints. 
Of particular interest is that ADM can accommodate very light dark matter, in the cold as well as in the warm regimes.

The scenario we have outlined may be varied in many ways.  
In this paper, we discuss two simple variations of the hidden sector in more detail. 
In one realization  the $\phi$ asymmetry, generated by $N_1$ decays, is {\em not} washed out during leptogenesis, but instead survives together with the $\chi$  asymmetry. 
We will see that the two asymmetries have about the same size, $n_{\Delta \chi} \simeq n_{\Delta \phi}$.  
Then, at later times, when annihilations are already too slow to significantly reduce the DM abundance,  $\phi$ decays to $\bar \chi$ (plus SM states) through an interaction generated by integrating out the right-handed neutrinos, 
and cancels out the DM asymmetry. 
Thus, in this scenario, the DM relic density is set by the primordial asymmetry, but today DM is symmetric.  
This fact has important consequence for phenomenology, notably for indirect detection of DM, as DM particles in our Galaxy may annihilate with a large cross-section.     

In another realization, $\phi$ obtains a VEV\@.   Consequently, through $N_i$ interactions, $\chi$ mixes with the left-handed neutrinos of the SM\@.   This scenario is a novel realization of the sterile neutrino framework for DM (for a review and references, see~\cite{Kusenko:2009up}), admitting a new and simple mechanism to populate its density.  The mixing with neutrinos opens up decay modes of DM into SM states.   
As we show below, for sufficiently small $\lambda \left< \phi \right>$, $\chi$ has a long enough lifetime to be consistent with current bounds, while the decays may be observable in the near future.  
The VEV of $\phi$ also generates a Majorana mass for $\chi$, leading to particle/antiparticle oscillations for DM, $\chi \leftrightarrow \bar \chi$.  We show that these oscillations can  lead to a large annihilation cross-section in our Galaxy.  
 
The paper is organized as follows.   
In Sec.~\ref{sec:exampleModel}, we explain the workings of 2-sector leptogenesis.  
To this end we construct a simple toy model that captures most of the physics  and discuss washout and transfer effects that influence the final asymmetries.
Sec.~\ref{sec:fullSM} describes the complete SM plus hidden sector scenario.
In  Sec.~\ref{sec:SDM}, we discuss the variations of our scenario in which, although the  DM relic density is set by the $\chi-\bar \chi$  asymmetry, we can repopulate symmetric DM at the present epoch. 
In  Sec.~\ref{sec:cosmo}, we discuss some additional constraints on our scenario that arise when the DM particle is lighter than the GeV scale. 
Our concluding discussion appears in Sec.~\ref{sec:conclusions}.   
In the Appendix we discuss some technical details of the Boltzmann equations of 2-sector leptogenesis.
We note that readers who are most interested in dark matter model building, and less familiar with the technicalities of leptogenesis, may prefer to read sections \ref{sec:toymodel}-\ref{sec:BEs} where we describe the toy model, and the discussion around Eqs.~\ref{eq:FinalAsymm1} and \ref{eq:FinalAsymm2}, where we discuss how the generated asymmetries relate to the DM mass.  It is then possible to skip to sections \ref{sec:SDM} and \ref{sec:cosmo}, where we discuss model building issues.  On the other hand, readers who are already familiar with leptogenesis may prefer to skip directly to section~\ref{sec:fullSM}, where we discuss the full model.

\section{Toy Model for Two-Sector Thermal Leptogenesis}
\label{sec:exampleModel}

In this section we discuss the simultaneous generation of matter-antimatter asymmetries in the SM and hidden sector during thermal leptogenesis.  
In order to highlight the relevant physics of 2-sector leptogenesis, we start with a simple toy model.  
The extension of the theory to accommodate the full SM is straightforward, and we highlight the ingredients in Sec.~\ref{sec:fullSM}.

\subsection{Toy Model}
\label{sec:toymodel}

Consider two fermion (matter) fields $l,\chi$ and two 
complex scalars $h,\phi$ coupled to two  Majorana neutrinos $N_i$, $i=1,2$,
\be
\label{eq:Lagrangian}
-\cL \supset \frac{1}{2} M_i N_i^2 + y_{i} N_i l h +   \lambda_i N_i \chi \phi + h.c. 
\ee
As the notation suggests, $l,h$ is a proxy for the SM sector, while $\chi,\phi$ represent the DM sector.
At low energy, the theory admits an approximate global ``lepton" symmetry under which $l$ and $\chi$ are charge $+1$ while $N_i$ are charged $-1$.  The symmetry is exact in the limit $M_i \to \infty$. 
On the other hand, we assume here that the quantum numbers carried by  $h$ and $\phi$ are rapidly washed out by some other interactions. 
We take both $\chi$ and $\phi$ to get masses at low energies, and DM stability follows from $m_\chi < m_\phi$, and an assumed ${\mathbb Z}_2$ symmetry under which both $\chi$ and $\phi$ are charged.
 The DM mass must be Dirac in order to preserve lepton number, so we assume that DM gets a mass with another fermion, $m_\chi \chi \tilde \chi$,  where $\tilde \chi$ has lepton number $-1$.
Furthermore, we assume the presence of additional lepton conserving interactions 
that rapidly thermalize $l,h,\chi,\phi $ and  ultimately annihilate the symmetric component $l + \bar l$, $\chi + \bar \chi$.\footnote{The canonical example for such  interactions is a $U(1)_d\times G_{\rm SM}$ gauge symmetry under which each sector is charged separately. 
Washout of  $h$ and $\phi$ asymmetries can be due to Yukawa interactions with other light fermions in the theory.
}
We will discuss these important model building issues later in this paper, but for the moment we focus on the mechanism generating asymmetries in the two sectors.
The key point of this example is to demonstrate that $l$  and $\chi$  can easily have different asymmetries, allowing for a large range of DM masses.
 
\subsection{Decay Asymmetry}

In order to generate an asymmetry in leptons and in DM, there must be $CP$-violation in the decays of $N_i$.  
While one Yukawa phase in each sector  can be removed by field redefinitions, 
the remaining 2 Yukawa phases are physical and lead to CP violation.  
We use the basis where $y_1$ and $\lambda_1$ are real and positive while $y_2=|y_2| e^{i \phi_\chi}$ and $\lambda_2 = |\lambda_2| e^{i \phi_l}$.  We take the hierarchal approximation, $M_1 \ll M_2$ and assume that we can integrate out $N_2$ and only include $N_1$ in the Boltzmann equations.  
We're interested in asymmetries in the decays of $N_1$,
\be
\label{eq:DefineAsymm}
\epsilon_\chi = \frac{\Gamma \left( N_1 \rightarrow \chi \phi  \right)-\Gamma \left( N_1 \rightarrow \bar \chi \phi^\dagger \right)}{\Gamma_{N_1}} 
\quad ,\quad \epsilon_l = \frac{\Gamma \left( N_1 \rightarrow l h \right)-\Gamma \left( N_1 \rightarrow \bar l h^\dagger \right)}{\Gamma_{N_1}}\,,
\ee
where $\Gamma_{N_1} = (y_1^2 + \lambda_1^2)M_{N_1}/16\pi$ is its total width.   
The asymmetries are straightforward to compute, 
\be
\label{eq:Asymm1}
\epsilon_\chi &\simeq& \frac{M_1}{M_2} \frac{1}{16 \pi (y_1^2+\lambda_1^2)} 
\left( 2 \lambda_1^2 |\lambda_2|^2 \sin \left( 2 \phi_\chi \right)+ y_1 y_2 \lambda_1 |\lambda_2| \sin \left(\phi_l + \phi_\chi \right) \right)  \,,
\\
\label{eq:Asymm2}
\epsilon_l &\simeq& \frac{M_1}{M_2} \frac{1}{16 \pi (y_1^2+\lambda_1^2)} 
\left( 2 y_1^2 |y_2|^2 \sin \left( 2 \phi_l \right) + y_1 y_2 \lambda_1 |\lambda_2| \sin \left(\phi_l+\phi_\chi \right)\right)\,.
\ee
We see immediately that the 2 sectors may have different asymmetries, and in particular, 
\be
\label{eq:AsymmRatio}
\frac{\epsilon_l}{\epsilon_\chi}\simeq \frac{2 r \sin (2 \phi_l)+\sin(\phi_l+\phi_\chi)}{2 r^{-1}\sin(2 \phi_\chi)+ \sin(\phi_l+\phi_\chi)}
\qquad , \qquad r = \frac{y_1 |y_2|}{\lambda_1 |\lambda_2|}\,.
\ee
Therefore $\epsilon_l / \epsilon_\chi \simeq r$ for generic phases. 
When the couplings of matter fields to both right handed neutrinos  are similar, $y_1 \simeq y_2$ and $\lambda_1 \simeq \lambda_2$, the asymmetry for each sector scales as the branching ratio of $N_1$ decays into that sector.  
Of course, when Yukawa couplings within one sector are hierarchical, e.g.  $y_1 \gg y_2$ and/or $\lambda_1 \ll \lambda_2$, 
the decay asymmetries do not have to be correlated with the branching ratios.

The final asymmetry in each sector is determined not only by the decay asymmetries $\epsilon_x$, but also by washout effects and transfer effects that may change the asymmetry in one or both of the sectors.  These  
may change the simple dependence on the branching ratios  quite drastically and may even result with a {\em larger} asymmetry in the sector with the {\em smaller} branching fraction and decay asymmetry.     
The range of possible asymmetry patterns is therefore very rich. 

\subsection{Boltzmann Equations}
\label{sec:BEs}

The cosmological evolution of the sterile neutrinos and the asymmetries are described by the Boltzmann Equations (BEs).  
We introduce the abundance yield $Y_{x} = n_{x}/s$ where $n_{x}$ is the number density of the particle $x$ and $s$ is the entropy density.
We are interested in the evolution of the asymmetries $Y_{\Delta l,\chi} = Y_{l,\chi} - Y_{\bar l,\bar \chi}$ as a function of time (or temperature $T$), assuming these asymmetries vanish at early times. 
To this end we solve the BEs that include the $N_1$ decays, inverse decays, and 2-to-2 scattering of matter in both sectors. 
For the initial conditions we assume that the matter in the two sectors is in equilibrium with the same temperature, while for $N_1$ we consider two cases: either equilibrium or zero abundance at early times. 
For the toy model at hand the BEs take the schematic form,
\begin{eqnarray}
\label{eq:BE}
\frac{s H_1}{z} Y_{N_1}' &=& -\gamma_D \left( \frac{Y_{N_1}}{Y_N^{\rm eq}} -1 \right) \, \,+ \, \, (2 \leftrightarrow 2) \,,
\\
\label{eq:BE2}
\frac{s H_1}{z} Y_{\Delta \chi}' &=& \gamma_D \left[ \epsilon_\chi \left( \frac{Y_{N_1}}{Y_{N_1}^{\rm eq}} -1 \right) - \frac{Y_{\Delta \chi}}{2 Y_\chi^{\rm eq}} \, \mathrm{\br}_\chi \right]  \, \,+ \, \, (2 \leftrightarrow 2 ~\textrm{washout + transfer}) \,,
\\
\label{eq:BE3}
\frac{s H_1}{z} Y_{\Delta l}' &=& \gamma_D \left[ \epsilon_l \left( \frac{Y_{N_1}}{Y_{N_1}^{\rm eq}} -1 \right) - \frac{Y_{\Delta l}}{2 Y_l^{\rm eq}} \, \mathrm{\br}_l\right] \, \,+\, \, (2 \leftrightarrow 2 ~ \textrm{washout + transfer}) \,.
\end{eqnarray}
Here $z = M_{N_1}/T$,  $H_1$ is the Hubble parameter at $T = M_{N_1}$, $s$ is the entropy density, $Y_{N_1,l,\chi}^{\rm eq}$ are the equilibrium number densities,
 and $\mathrm{\br}_{\chi,l}$ denote the branching fractions of $N_1$ into the two sectors.
Finally, $\gamma_D$ is  the thermally averaged $N_1$ decay density,
\be
\label{eq:GammaD}
\gamma_D = \frac{m_{N_1}^3  K_1(z)} {\pi^2 z}  \Gamma_{N_1}\,.
\ee
Further details and the complete set of equations are given in the Appendix.

\begin{figure}[t]
\begin{center}
\includegraphics[scale=0.5]{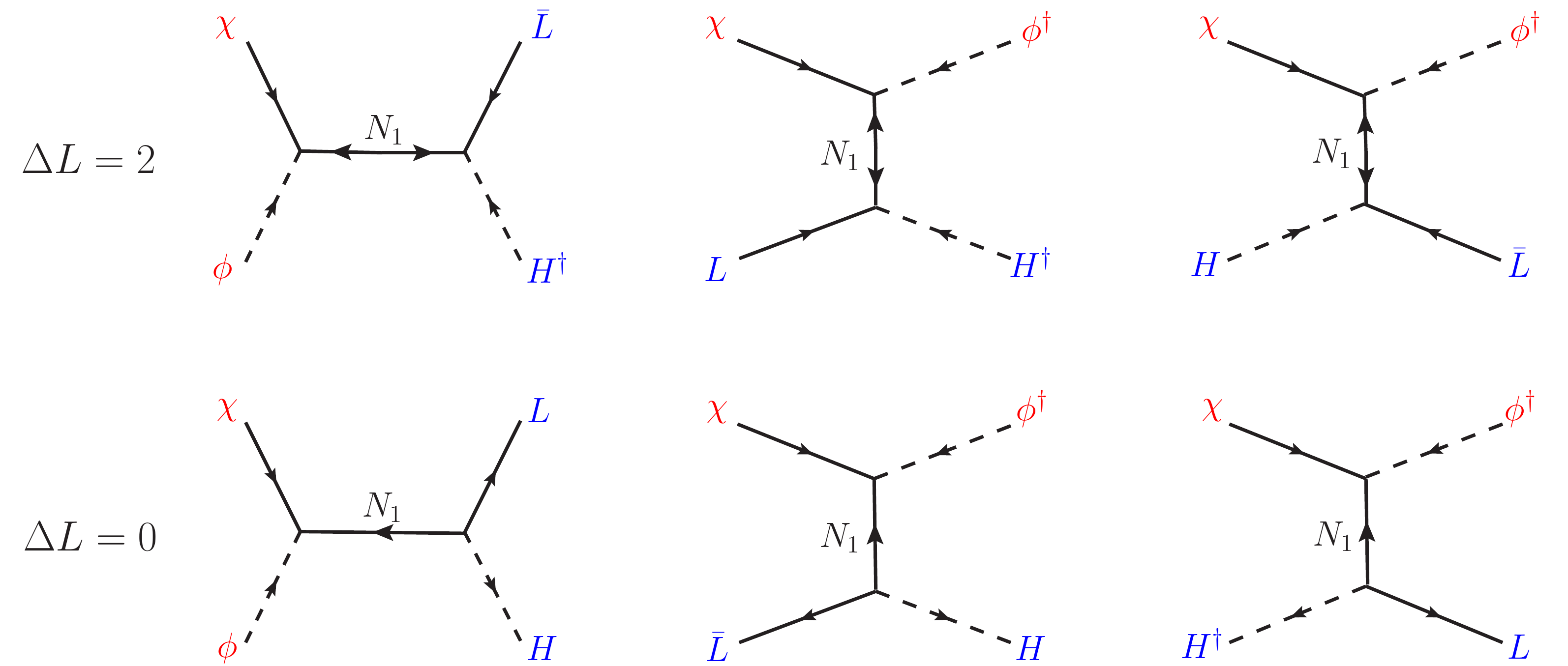}
\caption{\small\it  Feynman diagrams contributing to the 2-to-2 terms in the Boltzmann Eqs.~\eqref{eq:BE2} and \eqref{eq:BE3}, that transfer the lepton asymmetry between the two sectors.  The top row shows diagrams that violate lepton number, while the transfer diagrams in the bottom row conserve lepton number. }
\label{fig:transfer}
\end{center}
\end{figure}

The first equation describes the evolution of $N_1$ abundance due to decays and inverse decays. 
The strength of these interactions is set by $\gamma_D$ (or more appropriately $\Gamma_{N_1}/H_1$), which thus controls the departure of $N_1$ from thermal equilibrium. 
We will always assume that  $\Gamma_{N_1}/H_1$ is not too small, so that $N_1$ decays  before dominating the energy density of the universe.
In the other two equations,  the terms proportional to $\epsilon_{l,\chi}$ source the asymmetries $Y_{\Delta \chi, \Delta l}$ once $N_1$ drops out of thermal equilibrium. 
The  terms proportional to  ${\rm \br}_x$ describe the effect of $2 \to 1$ inverse decay processes $a a  \to N_1$  ($a  = l,\chi$ or the corresponding scalars) which lead to a washout of the asymmetries. 
The $(2\leftrightarrow 2$ washout$)$ stands for $\Delta L =2$ processes $a a \leftrightarrow \bar a \bar a $ with an off-shell $N_1$, while the ($2\leftrightarrow2$ transfer) terms stand for $\Delta L =0$ or 2 processes, $aa \leftrightarrow bb,\bar b \bar b$, that transfer the asymmetries from one sector to the other. 
 The Feynman diagrams contributing to these (2-to-2 transfer) terms, are shown in Fig.~\ref{fig:transfer}.

There are two basic regimes of the BEs (\ref{eq:BE}-\ref{eq:BE3})
\begin{itemize}
\item 
The {\bf narrow-width approximation}: $\Gamma_{N_1} \ll  M_{N_1}$ and  $\Gamma_{N_1}^2/ M_{N_1} H_1 \ll 1$. 
Here, the inverse decays are the dominant source of the washout and the 2-to-2 contribution can be neglected.  
Consequently, the last two equations decouple from each other and the asymmetries evolve independently in each sector.   
One can further distinguish the {\em weak washout} regime,  $\Gamma_{N_1} \ll  M_{N_1}$,  and the {\em strong washout} regime $\Gamma_{N_1} \gtrsim M_{N_1}$. In the latter, the asymmetries can be sizably reduced by the inverse decays.   
\item The {\bf large washout/transfer regime}: 
 $\Gamma_{N_1} \simeq  M_{N_1}$ or $\Gamma_{N_1}^2/ M_{N_1} H_1 \gtrsim 1$ . 
Here the 2-to-2 contributions are important and may change the final asymmetries by many orders of magnitude.
The last two BEs never decouple, and the 2 asymmetries get correlated due to strong washout and transfer effects.  
\end{itemize}

\begin{figure}[t]
\begin{center}
\includegraphics[width=6.6in]{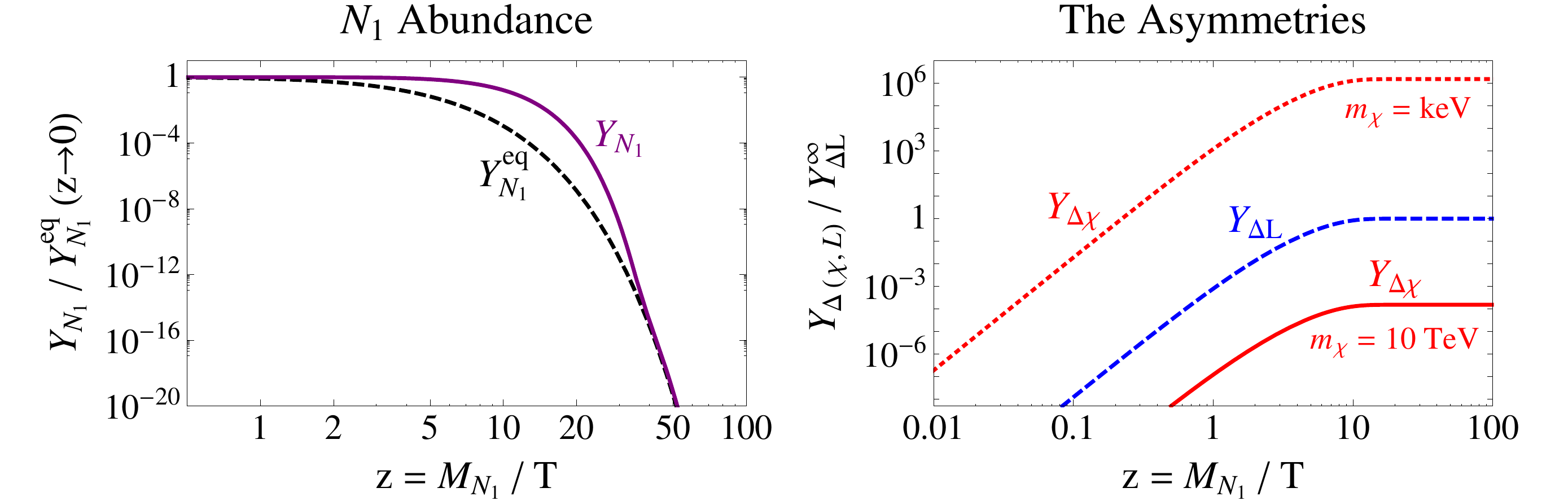}
\caption{\small\it Solutions to the Boltzmann equations for a 2-sector toy model with both sectors in the weak washout regime, $\Gamma_{N_1}\br_{l,\chi}/H_1\ll1$ (and consequently in the narrow-width limit), 
assuming the initial condition $Y_{N_1} = Y_{N_1}^{eq}$. 
 In this limit, the washout efficiencies are $\eta_{l,\chi}=1$ and the final lepton and DM abundances depend only on $\epsilon_{l,\chi}$ as in Eq.~\eqref{eq:eta}.   The {\bf left} plot shows the $N_1$ abundance (purple line) as a function of $z=M_{N_1}/T$, with its equilibrium value, $Y_{N_1}^{\rm eq}$, plotted for reference (black dashed).  The {\bf right} plot shows the asymmetry  abundances normalized to the asymptotic lepton abundance for ${\Delta L}$ (blue dashed) and ${\Delta\chi}$ with $m_\chi = $ keV (red dotted) and $m_\chi=10$ TeV (red line).}
\label{fig:weakweak}
\end{center}
\end{figure}

To describe these washout effects, it is  convenient to parametrize the asymptotic asymmetries, 
 \be \label{eq:eta}
 Y_{\Delta x}^\infty  = \eta_x \epsilon_x Y_{N_1}^{eq}(0) \,.
 \ee
In the narrow-width approximation, $Y_{\Delta x}^\infty $ is proportional to $\epsilon_x$ which parametrizes the CP-asymmetry of $N_1$ decays, as discussed earlier.
 In addition, the washout efficiencies, $\eta_x$, defined through the above parametrization, capture the washout effects occurring during the thermal evolution.   Note however that, while intuitive, these definitions need not imply that $N_1$ ever reaches it's equilibrium abundance.  
 In Sec.~\ref{sec:washout} we highlight some effects of the 2-to-2 scatterings.  We comment that in the narrow-width limit, since the BEs for the two asymmetries are decoupled, the efficiencies are bounded, $\eta_x <1$.  However, more generally, we will see that 2-to-2 transfer effects can dominate one of the asymmetries, leading to $\eta_x>1$.
For now, we show in Fig.~\ref{fig:weakweak} simple solutions to the BEs for which  $\eta_{L,\chi} = 1$, occurring in the narrow-width limit and the weak-washout regime $\Gamma_{N_1} / H_1 \ll 1$.  The solutions assume an initial thermal abundance of $N_1$ and demonstrates the two viable DM mass limits of keV and $\sim10$ TeV\@.

\subsection{Washout Effects}
\label{sec:washout}

Washout effects may play an important role in one or both sectors.  
 For a given sector, one can distinguish between weak, $\br_x \Gamma_{N_1}  /H_1 \ll 1$, and strong, $\br_x \Gamma_{N_1}/H_1\gg 1$ washout.
 We now briefly highlight a few interesting washout effects.   
To classify the spectrum of possibilities it is convenient to divide the discussion into three cases depending  on whether each sector is in the weak or strong washout regime. 

\subsubsection{Weak/Weak}

In this case $\br_{L,\chi} \Gamma_{N_1}/H_1\ll 1$.  It follows that for $M_{N_1}\lesssim 10^{18}$ GeV,  the two sectors are in the narrow-width regime, $\br_{L,\chi} \Gamma_{N_1}/M_{N_1}\ll 1$, and are therefore decoupled.  Consequently, washout is negligible and the final asymmetry strongly depends on whether $N_1$ thermalizes before decaying.  In the case that it does the final asymmetries are set by $\epsilon_x$ and one has,
\be
\eta_L \simeq 1\ , \qquad \eta_\chi \simeq 1 \,.
\ee
This situation is depicted in Fig.~\ref{fig:weakweak}.   The  ratio of the asymmetries,  
\be
R_{\Delta} \equiv Y_{\Delta L}^\infty /Y_{\Delta \chi}^\infty\,,
\ee
 is then simply  $R_{\Delta} \simeq \epsilon_{L}/\epsilon_{\chi}$, which can be extracted from Eqs~\eqref{eq:Asymm1} and~\eqref{eq:Asymm2}.  
In particular, for our toy model, $R_{\Delta} \simeq  y_1 y_2/\lambda_1 \lambda_2$.
A hierarchical $R_\Delta$ may appear if the Yukawa couplings display a hierarchy between the 2 sectors.

\begin{figure}[t]
\begin{center}
\hspace{-.7cm}
\includegraphics[scale=0.6]{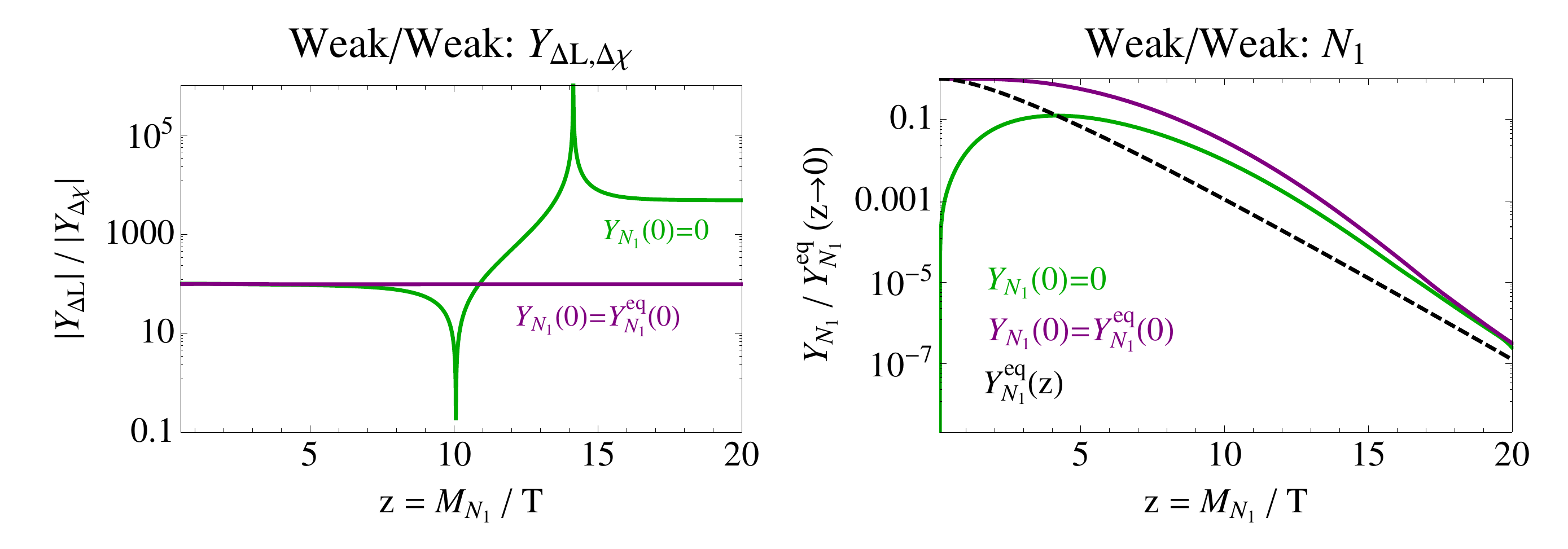}
\caption{\it\small Solutions to the Boltzmann equations in the case where both sectors are in the weak washout regime $\Gamma_{N_1}\br_{L,\chi}/H_1\ll1$ (which implies the narrow-width approximation),  
for $\br_\chi = 10^{-2}$ and $\epsilon_{L,\chi} = 10^{-5}\times \br_{L,\chi}$. 
Two solutions are shown, assuming thermal and zero initial conditions for $N_1$.  The former implies washout efficiency of order one, $\eta_{L,\chi}=1$ as in Fig.~\ref{fig:weakweak}.  On the other hand, in the latter case $N_1$ never thermalizes and consequently the efficiencies are  smaller, as predicted in Eq.~\eqref{eq:weakweakeff}.  The {\bf left} plot shows the ratio of lepton to dark matter abundance as a function of $z=M_{N_1}/T$ while the {\bf right} plot shows the normalized $N_1$ abundance.    }
\label{fig:weakweak2}
\end{center}
\end{figure}

If $Y_{N_1}(0) = 0$,  the asymmetry vanishes in the first approximation, as a negative asymmetry generated at $z \ll 1$ (when $Y_{N_1}$ is still less than $Y_{N_1}^{eq}$ ) cancels against the positive asymmetry generated at $z > 1$ (when $Y_{N_1} > Y_{N_1}^{eq}$). 
A small asymmetry arises thanks to the washout effects being different at small and at large $z$.  One can estimate, 
\be
\label{eq:weakweakeff}
\eta_L \simeq \frac{\Gamma_{N_1}^2}{H_1^2}\br_L \,,
\qquad 
\eta_\chi \simeq \frac{\Gamma_{N_1}^2}{H_1^2}\br_\chi \,.
\ee
Thus  $R_\Delta \simeq \epsilon_{L} {\rm \br}_{L}/\epsilon_{\chi} {\rm \br}_{\chi}$, which, again for the toy model, gives $y_1^3 y_2/\lambda_1^3 \lambda_2$. 
In this case even a small hierarchy between  the Yukawa couplings of  the 2 sectors  to heavy neutrinos  may be amplified into a hierarchical $R_\Delta$.   
In Fig.~\ref{fig:weakweak2} we demonstrate the above scaling with a solution in the weak/weak limit for both $Y_{N_1}(0) = 0$ and $Y_{N_1}(0) = Y_{N_1}^{\rm eq}(0)$.  

\subsubsection{Strong/Strong}
\label{sec:strongstrong}

\begin{figure}[t]
\begin{center}
\includegraphics[width=6.6in]{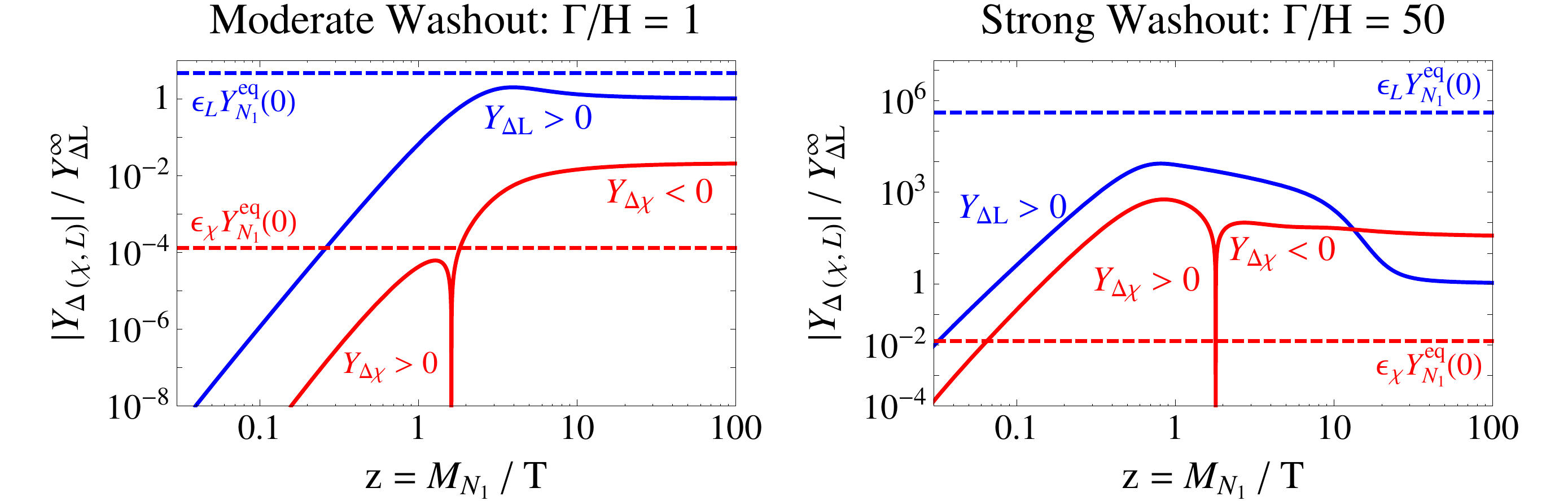}
\caption{\it\small Normalized abundances of lepton and DM asymmetries as a function of $z=M_{N_1}/T$.  The dashed curves show the expected asymptotic asymmetries for unit washout efficiencies, $\eta_{L,\chi}=1$.  The {\bf left} plot shows the solution in the case $\Gamma_{N_1}/H_1 =1$.  The DM asymmetry changes sign due to the significant washout and transfer effects.  The final asymmetry in that sector is greater than one.  Here $\Gamma_{N_1}/M_{N_1} = 0.1$ and $\br_L=0.9$.   On the {\bf right} plot, the solution is shown with identical parameters except for $\Gamma_{N_1}/H_1=50$.  The corresponding theory is in the strong/strong regime with wide $N_1$ width.  As can be seen, the large washout and transfer effects reverse the ratio of lepton to DM abundance, rendering a larger number density in the dark sector.  As discussed in Sec.~\ref{sec:washout} the ratio of the asymptotic abundances is independent of $\epsilon_{L,\chi}$.  }
\label{fig:strongstrong}
\end{center}
\end{figure}

For large $\br_{L,\chi} \Gamma_{N_1}/H_{1}\gg 1$ there is no significant dependence on initial conditions for $N_1$, 
however there is a qualitative dependence on whether the narrow-width approximation holds.  
For $\Gamma_{N_1} \ll M_{N_1}$  and $\Gamma_{N_1}^2 \ll  M_{N_1} H_1$ transfer and $2\leftrightarrow 2$ washout effects are suppressed, and the asymmetry in each sector evolves independently. 
One can estimate the washout efficiencies as  
\be
\label{dupa1}
\eta_L \simeq \frac{H_1}{\Gamma_{N_1}} \frac{1}{\br_L}\ , \qquad \eta_\chi\simeq \frac{H_1}{\Gamma_{N_1}} \frac{1}{\br_\chi}\,.
\ee
The asymmetry ratio now scales as $R_\Delta\simeq  \epsilon_{L}{\rm \br}_{\chi}/\epsilon_{\chi} {\br}_{L} \simeq \lambda_1 y_2/y_1 \lambda_2$.
This is of order one unless the couplings to $N_1$ and $N_2$ display a hierarchy in at least one of the sectors.
Thus, assuming no hierarchy of Yukawa couplings within each sector in this case, one predicts comparable asymmetries in the two sectors. 
In the ADM scenario that would translate to  a prediction that the DM mass is comparable to the baryon mass. 
 Thus the strong-strong case  without Yukawa hierarchies may be linked to  the  ADM scenarios in the literature that predict $Y_b/Y_{DM} \sim 1$.
 In reality, however, the Yukawa couplings in the SM sector typically do display a hierarchy to match the observed neutrino masses, which  destroys this prediction.  

For $\Gamma_{N_1} \simeq M_{N_1}$ and/or $\Gamma_{N_1}^2 \ll  M_{N_1} H_1$ washout gets amplified  due to the contribution of 2-to-2 processes,  
which may change the efficiency estimates in  Eq. (\ref{dupa1}) by many orders of magnitude. See the Appendix for details. 
The asymmetries in the two sectors can then get correlated due to transfer effects and since the transfer terms in Eqs.~\eqref{eq:BE2},~\eqref{eq:BE3} are proportional to the branching fraction, one expects $R_\Delta  \propto \br_\chi/\br_L$ independently of $\epsilon_x$.  Interestingly, these results imply that far from the narrow-width limit, it is generic to have a {\it larger} density in the sector with {\it smaller} branching fraction.    
This is demonstrated on the right side of Fig~\ref{fig:strongstrong}:  $R_\Delta$ is indeed found to be  independent of $\epsilon_x$ and the density in the DM sector dominates.   
One can also note that the DM density changes sign due to the strong transfer effects.

\subsubsection{Strong/Weak}

An intermediate case occurs when only one sector (say the SM), is in the strong washout regime while the other (here the DM sector)  is in the weak washout regime.  
In the narrow-width approximation one then finds 
\be
 \eta_L \simeq \frac{H_1}{\Gamma_{N_1} \br_L}\,,  \qquad 
\eta_\chi \simeq  \left\{\begin{array}{lll} 1 && Y_{N_1}(0) =Y_{N_1}^{eq} \\ \frac{\Gamma_{N_1}}{H_1}\br_\chi&&Y_{N_1}(0) = 0\end{array}\right. \,.
\ee
In the first case $R_\Delta$ is suppressed  from the naive value $\epsilon_L/\epsilon_\chi$ by a small factor $H_1/\Gamma_{N_1}\br_L$  (while in the other case the factor is  $H_1^2/\Gamma_{N_1}^2 \br_L \br_\chi$ which may or may not be small). 
As in the strong/strong case, the above suppression may allow for a situation where the density is larger in the sector with smaller branching fraction.  For instance, for $\epsilon_L/\epsilon_\chi=1$, $\Gamma_{N_1}/H_1=10$ and $\br_\chi= 10^{-2}$, one finds an order of magnitude larger density in the hidden sector.
For  $\Gamma_{N_1} \ll M_{N_1}$  and $\Gamma_{N_1}^2 \ll  M_{N_1} H_1$   the washout of $Y_{\Delta l}$ becomes even larger, further strengthening the aforementioned effect, while loosing the $\epsilon_x$ dependence of the ratio.

\section{Towards a Complete Model}
\label{sec:fullSM}

Let us now briefly discuss how the above model is modified when we replace one of  the sectors with the complete SM, 
\be
\label{eq:LagrangianQM}
-\cL \supset \frac{1}{2} M_i N_i^2 + Y_{i \alpha} N_i L_\alpha H  +   \lambda_i N_i \chi \phi  +  h.c.  \,,
\ee
where $\alpha = 1 \dots 3$ counts the SM generation.
As in in the toy model, we assume that DM pairs up with another fermion, $\tilde \chi$, to receive a Dirac mass at low energies, $m_\chi \chi \tilde \chi$. 
We define the decay asymmetry into the SM as the sum of decay asymmetries into each generation, $\epsilon_L = \sum_\alpha \epsilon_{L_\alpha}$.  
$\chi$ is now the asymmetric DM candidate.  
To match observation, the asymptotic asymmetries should have the numerical values,
\be
Y_{\Delta L}^\infty = \epsilon_L \eta_L \, Y_{N_1}^{eq} \left( 0 \right)& \simeq &  2.6 \times10^{-10}  \label{eq:FinalAsymm1} \\
Y_{\Delta \chi}^\infty = \epsilon_\chi  \eta_\chi \, Y_{N_1}^{eq} \left(0\right) & \simeq & 4 \times 10^{-10} \left( {\rm GeV} \over m_\chi \right) \label{eq:FinalAsymm2}
\ee
where $Y_{N_1}^{eq} \, (0) = 135 \zeta(3) / 4 \pi^4 g_*$ and  $g_* \sim  100 $ counts the total number of relativistic degrees of freedom at $T  \sim M_{N_1}$.  
On the right-hand side, we show the asymmetries that are required to reproduce the observed baryon and DM abundances, assuming $Y_{B} = 12 Y_{\Delta L}/37$.   

The value for $\epsilon_L \eta_L$ is fixed by experiment (up to a small dependence on $g_*$ at $T \sim M_{N_1}$).
 On the other hand, a prediction for $\epsilon_\chi \eta_\chi$ translates into a prediction for the DM mass required to match the observed abundance. 
For example, if the set-up predicts  $Y_{\Delta l}^\infty /Y_{\Delta \chi}^ \infty \sim 1$, the required DM mass is in the GeV ballpark. 
However, as we discussed in the toy model,  such a relation between the asymmetries is not a generic prediction of two-sector leptogenesis in most of the parameter space:    
the ratio of the decay asymmetries  $\epsilon_L/\epsilon_{\chi}$ depends on arbitrary Yukawa couplings, and moreover there is a  wide spectrum of possible washout efficiencies $\eta_{L,\chi}$.  
All this implies that a large range of dark matter masses are possible.

\begin{figure}[t]
\begin{center}
\includegraphics[width=6.6in]{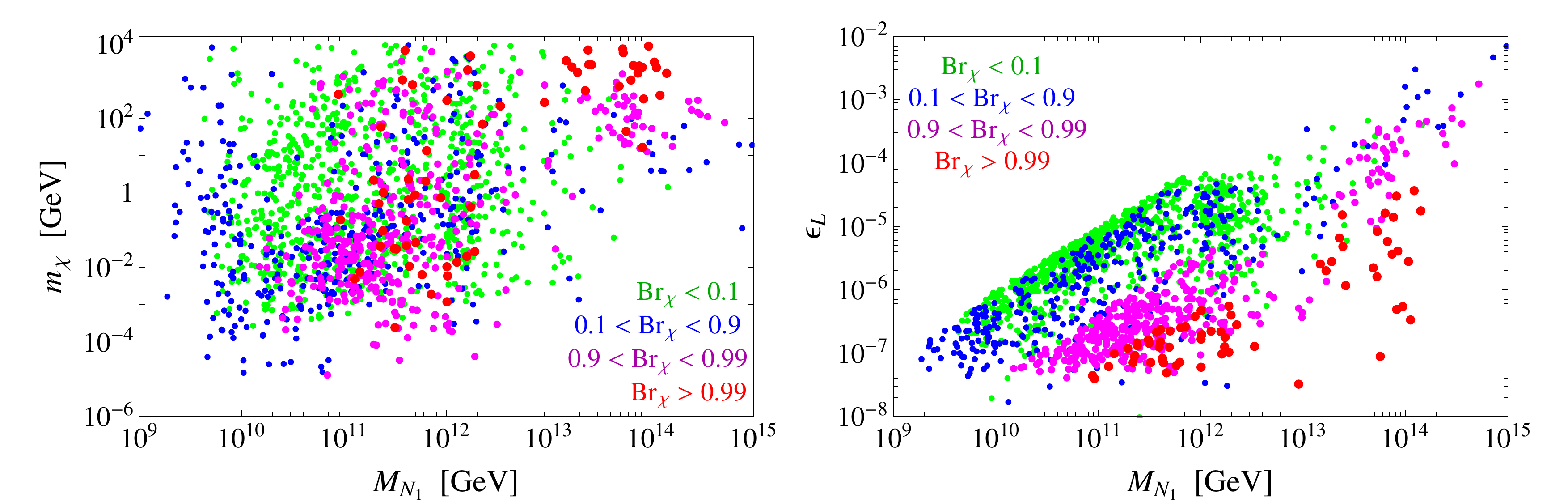}
\caption{\small\it  Scatter plots for realistic 2-sector thermal leptogenesis, scanning over models that generate the correct SM lepton asymmetry, Eq.~\eqref{eq:FinalAsymm1}, while at the same time producing the correct spectrum and mixing angles for the active neutrinos.  The {\bf left} plot shows the spread of DM masses as a function of the mass of the lightest right-handed neutrino, assuming $M_{N_3}/M_{N_2} = M_{N_2}/M_{N_1} = 10$.  The coloring of points indicate the branching fraction of $N_1$ into the hidden sector.   10 keV to 10 TeV masses are accommodated within the thermal leptogenesis framework.  Very light DM is typically obtained for smaller $M_{N_1}$ and for hidden sector branching fractions of order $0.1-0.9$.  The {\bf right} plot demonstrates  the attainable values for the CP violating parameter, $\epsilon_L$, as a function of the lightest right-handed neutrino mass.   
For a given mass, $M_{N_1}$, a maximal  value for $\epsilon_L$ is clearly visible, in accordance with the DI bound.  For both plots, we assume that $N_1$ starts with a thermal abundance.}
\label{fig:cremebrulee}
\end{center}
\end{figure}

If the SM neutrino masses are generated through the see-saw mechanism, a generalized Davidson-Ibarra (DI) bound~\cite{Davidson:2002qv} on $M_{N_1}$ can be derived.   
Working in the hierarchical limit, $M_{N_1}\ll M_{N_{2,3}}$, we can express $\epsilon_{L,\chi}$ as,
\begin{eqnarray}
\label{eq:epsilonSM}
\epsilon_L &\simeq& \frac{M_{N_1}}{8\pi} \frac{{\rm Im}[(3Y^*Y^T + \lambda^*\lambda) M^{-1} YY^\dagger]_{11}}{[2YY^\dagger + \lambda\lambda^*]_{11}}\,,
\\
\label{eq:epsilonDM}
\epsilon_\chi &\simeq& \frac{M_{N_1}}{8\pi} \frac{{\rm Im}[(Y^*Y^T + \lambda^*\lambda) M^{-1} \lambda\lambda^*]_{11}}{[2YY^\dagger + \lambda\lambda^*]_{11}}\,,
\end{eqnarray}
where above $M = \textrm{diag}(M_{N_1},M_{N_2},M_{N_3})$.  Concentrating on $\epsilon_L$ and using the relation $[Y Y^\dagger]_{ij} = M_{N_i}^{1/2}M_{N_j}^{1/2}[R\ m_\nu\ R^\dagger]_{ij} / v_{EW}^2$ with $R$ an arbitrary orthogonal complex matrix, an upper bound is found,
\begin{equation}
\label{eq:DI2}
\epsilon_L  \leq  \frac{3 M_{N_1} m_{\nu}^{\rm max}}{16 \pi v_{\rm EW}^2} \ C 
\simeq  10^{-7}  \left(\frac{M_{N_1}}{10^{9} \textrm{ GeV}}\right)  \ C \,. 
 \end{equation}
Here $v_{\rm EW} = 174$ GeV is the VEV of the SM Higgs and  $m_\nu^{\rm max}$ is the heaviest neutrino which was taken to be $0.05$ eV on the RHS\@.  $C$ is a function of $M$ and the Yukawa matrices.  It is simply expressed in the limit where the $N_1$ branching fraction into one of the sectors dominates\footnote{Here ${\rm \br_L}$ denotes the total branching ratio of $N_1$ into the sum of the SM flavors.},
\begin{equation}
\label{eq:C}
C \simeq \left\{\begin{array}{lll} 1 && {\rm \br_L} \gg {\rm \br_\chi} \\ (\lambda_2^2M_{N_1}/\lambda_1^2M_{N_2})^{1/2} &&  {\rm \br_L} \ll {\rm \br_\chi}\end{array}\right.
\end{equation}
In deriving the small ${\rm \br_L}$ limit above, we assumed $R$ is a matrix with order one coefficients, to avoid tuning or non-perturbative Yukawa couplings.  We see that the standard DI bound is recovered when $N_1$ decays mostly to the SM\@.  
A large enough asymmetry in the SM requires $\epsilon_L \gtrsim 10^{-7}$ which then implies $M_{N_1} \gtrsim 10^{9}$ GeV\@. 
 In the opposite limit the DI bound is multiplied by a factor depending on the ratio of the Yukawa couplings in the dark sector, and on the ratio of the sterile neutrino masses.   
 This typically leads to an  additional suppression and the bound becomes stronger, 
 unless there is a large hierarchy $|\lambda_{2}/\lambda_1| \gg M_{N_2}/M_{N_1}$ accompanied by 
 $|\lambda_1| > |Y_{1 \alpha}| \gtrsim 10^{-2} (M_{N_1}/10^{9} {\rm GeV})^{1/2}$ . 
 The modified DI bound is clearly visible in the scatter plot in Fig. \ref{fig:cremebrulee}. 

An additional consequence of thermal leptogenesis with see-saw masses is that the SM sector typically lies in the strong washout regime, 
${\br_L} \Gamma_{N_1} /H_1 \gg 1$.   Indeed, one finds, 
\begin{equation}
\label{eq:SMwashout}
  \frac{ \br_L \Gamma_{N_1}}{H_1} 
=  \frac{M_{\rm Pl} }{ \sqrt{g_*/90}}\frac{[R \, m_\nu \,  R^\dagger]_{11} }{ 8\pi^2 v_{\rm EW}^2} 
 \simeq  25  {m_\nu^{\rm max} \over 0.05 ~{\rm eV} } \,,
\end{equation}
where $g_* \sim 100$ and  a generic $R$ was assumed (weaker washout  may however arise for $R_{12}, R_{13} \ll 1$).
The hidden sector, on the other hand, may lie either in the weak washout regime  for $\br_\chi \ll 1$,  or in the strong washout regime for $\br_\chi \sim 1$. 
Note that  $\br_L \ll 1$ implies that washout in the hidden sector becomes extremely strong.  
The multiplicity of available scenarios thus allows for a wide range of the asymmetries that can be generated.
In particular, it is straightforward  to make the dark sector asymmetry subdominant $Y_{\Delta \chi}^\infty \ll Y_{\Delta L}^\infty$, corresponding to DM mass larger than GeV\@.  For this the hidden Yukawa couplings have to be small, in which case, 
 $\epsilon_\chi$ is suppressed.
A theoretical bound on the DM mass,  $m_{\chi} \lesssim 10$~TeV, then arises from the perturbativity  bound on the annihilation rate of the symmetric DM component.\footnote{However larger DM masses may be possible when dark matter is composite and there is a hierarchy between the confining scale and the DM mass \cite{Arvanitaki:2005fa}.}

A lower limit on the DM mass in the 2-sector thermal leptogenesis scenario follows from perturbativity, which requires $\epsilon_\chi \lesssim 10^{-1}$. 
Since $Y_{N_1}^{eq} \simeq 4\times 10^{-3}$, it follows $Y_{\Delta\chi}^{\infty}\lesssim 4\times10^{-4}$ and  Eq.~\ref{eq:FinalAsymm2} yields the lower bound of $m_\chi  \gtrsim  $ keV\@.   
Coincidently, the rough astrophysical bound on hot DM is also of order keV \cite{Viel:2005qj}.
In reality, however, keV DM mass is hard to obtain in thermal leptogenesis since washout effects  typically suppress the initial production of the DM asymmetry even if the branching ratio into the hidden sector is large.  
 In principle, such washout effects can be suppressed if the two sectors are in the weak-washout regime  or if the branching ratio into the SM sector is large while at the same time $\epsilon_L$ is small.  
 Both of these possibilities are harder (but not impossible) to realize in our thermal scenario with see-saw neutrino masses, but can easily be found in deformations of this setup.  
In Fig.~\ref{fig:cremebrulee} we show scatter plots that demonstrate the mass reach in the thermal leptogenesis case.
In these scans we assume hierarchical sterile neutrino masses, $M_{N_1}:M_{N_2}:M_{N_3}=1:10:100$, which implies  $\epsilon_\chi \lesssim 10^{-2}$ and thus a slightly larger lower bound $m_\chi  \gtrsim  10$ keV\@.   
We see that the lower reach of $m_\chi$ is indeed roughly $10$ keV\@.   

Before closing this subsection, a few remarks are in order.   
\begin{itemize}
\item Throughout the paper we have ignored finite-temperature effects which may play a significant role in some corners of the parameters space~\cite{Giudice:2003jh}.  Nonetheless, we don't expect the conclusions to change qualitatively. 
\item The conclusions and plots in the above discussion rely strongly on the thermal leptogenesis scenario with neutrino masses arising from the see-saw mechanism.  
It is straightforward to consider other, less limiting scenarios.   
For instance the assumed hierarchy of the sterile neutrino masses does not need to exist, additional Higgs fields may be present,
or other  leptogenesis scenarios
can be the dominant source for the asymmetry.  In such cases, the DI bound takes a different form and the concluded possible DM mass spectrum may be very different.  
\item Adding the hidden sector is not exactly the same as adding an additional flavor, as the hidden sector comes with a new  scalar field. 
Nevertheless, many effects present in the context of 3-flavor leptogenesis~\cite{Abada:2006fw} are valid in this case too. 
\item In the above we integrated out $N_{2,3}$, ignoring the asymmetry produced from their decays. 
 As in the SM case, this is not always justified and special care may be needed if the Yukawa couplings in one sector display a large hierarchy~\cite{Engelhard:2006yg}.
\end{itemize}

\section{Symmetric Dark Matter from an Asymmetry}\label{sec:SDM}

In this section, we consider several simple variations of the framework introduced above.  In sections~\ref{sec:exampleModel} and~\ref{sec:fullSM}, we described the class of models where the decays of a right-handed neutrino, $N_1$, result in a dark matter asymmetry through the operator $N_1 \chi \phi$, where $\phi$ is a scalar belonging to the dark matter sector, taken to satisfy $m_{\phi} > m_{\chi}$.  We assumed that $\phi$ carries no asymmetry due to the presence of fast interactions that convert $\phi \leftrightarrow \bar \phi$.  We further assumed that $\phi$ does not receive a VEV, so that DM is stable.  In this section, we relax these two assumptions and consider models where:

\begin{enumerate}
\item {\bf An asymmetry for $\phi$ is generated}.
\item {\bf $\phi$ obtains a VEV}.
\end{enumerate}

We will see below that these simple modifications have important consequences for dark matter phenomenology.  In models of type (1), where $\phi$ also carries an asymmetry, the decays of $\phi$ to $\bar \chi$, at low temperature, will reintroduce symmetric dark matter at late times.  As we discuss in section~\ref{sec:PhiDecay}, models of this type predict a large DM annihilation rate in the present day.  In models of type (2), the VEVs of $\left <\phi \right> $ and the SM higgs, $\left< h \right>$, cause DM to inherent a small mixing with neutrinos.  In a sense,   DM becomes a sterile neutrino with a large Dirac mass, $m_\chi \chi \tilde \chi$.  This has two important phenomenological consequences which we discuss in section~\ref{sec:sterile}: DM can decay into SM fermions giving observable signatures, and $\chi$ can oscillate into $\chit$ at late times, also repopulating symmetric dark matter.   

In both cases discussed here, the symmetric DM component is obtained at late times from the dominating asymmetric one.   Of course, symmetric DM is also produced directly through $N_1$ decays and may dominate the energy density already at earlier times.  The predictions of such a scenario are distinct from the ones considered here, and will be presented in future work~\cite{futurework}.

\subsection{Restoring Symmetric Dark Matter with Late Decays}\label{sec:PhiDecay}

Above, in our example model of thermal leptogenesis, we assumed that $\phi$ has interactions that are efficient at low energy and set $n_\phi$ = $n_{\phi^\dagger}$.  This is roughly the situation for the SM Higgs, but does not need to hold in the hidden sector.  Consider the case where $\phi$ number is preserved, so that $n_{\Delta \phi} = n_{\phi} - n_{\phi^\dagger}$ can be nonzero at low temperatures.  As we discuss below, the asymmetry in $\phi$ results in the restoration of $\bar \chi$ after $\chi$ decouples.  This leads to symmetric DM with some interesting phenomenological possibilities that we discuss in this section: a large annihilation rate at the present day and mixed warm/cold dark matter.

\begin{figure}[t]
\begin{center}
\includegraphics[width=5.5in]{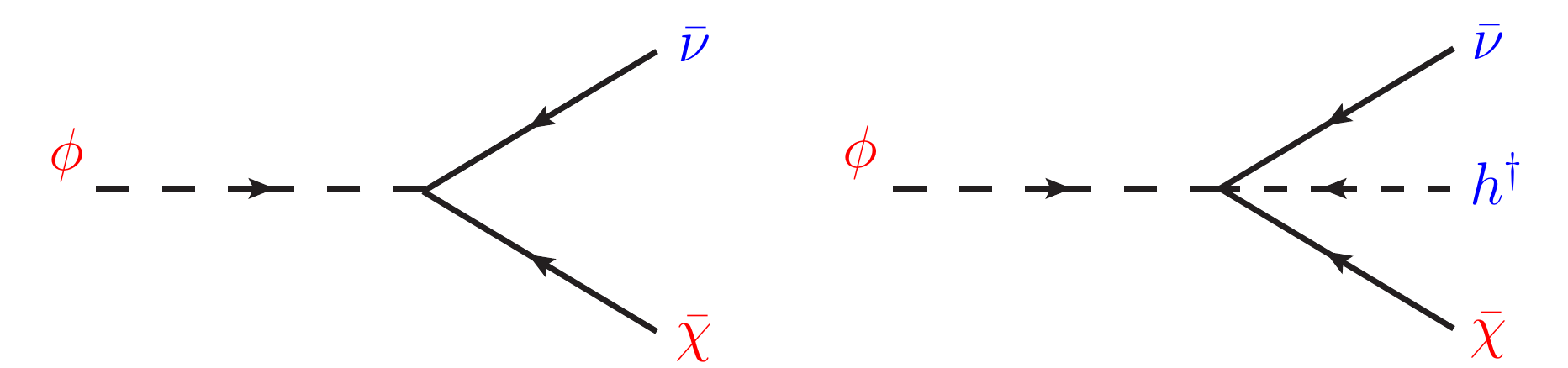}
\caption{\it\small Two-body and three-body decay modes of $\phi$ into DM and SM particles.  In the absence of washout processes for $\phi$, its asymmetry is equal to that of $\chi$.  Consequently, the above decays, which occur after annihilations have decoupled, repopulate the symmetric component of DM, allowing for an observable annihilation signal.}
\label{fig:phidecay}
\end{center}
\end{figure}

We now outline the cosmology of this scenario.  Suppose, as above, that CP violation in right-handed neutrino decays produce a $\chi$ asymmetry, $n_{\Delta \chi} >0$.   In the absence of strong washout processes, there will also be an asymmetry of $\phi$, $n_{\Delta \phi} > 0$.  In fact, if the BEs for the two asymmetries are invariant under the exchange of $\chi$ and $\phi$, the resulting asymmetries are equal:
\begin{equation}
\label{eq:DeltaPhiDeltaChi}
n_{\Delta \phi} =n_{\Delta \chi}\,.
\end{equation}
One can then check that a sufficient condition for the exchange symmetry to exist, is for the low energy interactions to preserve a $U(1)_L\times U(1)_\phi$ symmetry under which the fields transform as $\phi(0,1)$ and $\chi(1,-1)$\footnote{Here we assume  a Maxwell-Boltzmann distribution for the equilibrium values of the two fields.}.   These symmetries further guarantee that $n_{\Delta\chit} = 0$.  Incidentally the mass term, $m_\chi \chi\chit$ violates the above condition but keeps Eq.~\eqref{eq:DeltaPhiDeltaChi} intact.  If, however, other interactions exist that involve $\chit$, Eq.~\eqref{eq:DeltaPhiDeltaChi} is modified by an order one amount, reflecting the redistribution of the $\chi$ asymmetry into that of $\chit$.  Assuming no such interactions, when the temperature reaches the $\chi$ and $\phi$ masses, the symmetric components annihilate and freezeout, leaving asymmetric abundances of $\chi$ and $\phi$.  Then, at temperatures below $\chi$ and $\phi$ decoupling, the scalar $\phi$ can decay to $\bar \chi$ as shown in Fig.~\ref{fig:phidecay}, repopulating symmetric DM.  Eq.~\ref{eq:DeltaPhiDeltaChi} implies that the final abundances of $\chi$ and $\bar \chi$ are equal.

The Boltzmann equation for $\Delta \chi$ now also depends on $\Delta \phi$, but given Eq.~\ref{eq:DeltaPhiDeltaChi}, it can be expressed in a very similar form to equation Eq.~\ref{eq:BE2},
\be
 \frac{s H_1 }{z} Y_{\Delta \chi}' &=& \gamma_D \left[ \epsilon_\chi \left( \frac{Y_{N_1}}{Y_{N_1}^{eq}} -1 \right) - \frac{Y_{\Delta \chi}}{Y_\chi^{eq}} \, \br_\chi \right]  \, \,+ \, \, (2\leftrightarrow 2 ~\mathrm{washout + transfer}) \,.
\ee
The only difference, from above, is that the $2 \rightarrow 1$ washout term is twice as big as when $\phi$ had no asymmetry.

For simplicity, in the rest of this section, we assume the hierarchical limit $M_{N_1} \ll M_{N_{2,3}}$, and only include the lightest right-handed neutrino, $N_1$.  Furthermore, we suppress lepton flavor indices.  It is straightforward to extend our discussion to the more general case. 
In our example model, $\phi$ can decay to $\bar \chi$ through a dimension 5 operator that is generated by integrating out the right-handed neutrinos,
\be \label{eq:DecayOp}
\cL  \supset - y  \lambda \frac{\chi \phi L H}{M_{N_1}} + h.c.\,.
\ee
 Inserting the Higgs VEV, $\phi$ can two-body decay to $\bar \chi$ and a neutrino,
\be
\Gamma (\phi \rightarrow \bar \chi \bar \nu) = \frac{y^2 {\lambda}^2}{32 \pi} \frac{v_{\rm EW}^2}{M_{N_1}^2} m_\phi \left( 1 - \frac{m_\chi^2}{m_\phi^2} \right)^2 = \frac{\lambda^2}{16 \pi}  \frac{m_{\nu}}{M_{N_1}} m_\phi \left( 1 - \frac{m_\chi^2}{m_\phi^2} \right)^2,
\ee
where $m_{\nu}=y^2 v_{\rm EW}^2 / M_{N_1}$ is the see-saw contribution to the neutrino mass matrix, along the diagonal in flavor space, from integrating out $N_1$.  
Decays to neutrinos are only weakly constrained by Big Bang Nucleosynthesis (BBN), and decays as late as $\tau \lesssim 10^6\unit{sec}$ are allowed if $\phi$ decays entirely through this mode~\cite{Kanzaki:2007pd}.  Decays can easily proceed fast enough,
\be
\tau_{\phi} \simeq 7\times 10^{-4}\unit{sec} \times \left(\frac{0.1}{\lambda} \right)^2 \left(\frac{0.05\unit{eV}}{m_{\nu}} \right) \left( \frac{M_{N_1}}{10^{9}\unit{GeV}} \right) \left( \frac{100\unit{GeV}}{m_\phi} \right)\,.
\ee

Equation~\ref{eq:DecayOp} also introduces a three-body decay mode where the Higgs is produced, $\phi \rightarrow \bar \chi \nu h$.  This decay is of course suppressed by three-body phase space, but is enhanced, relative to the two body decay, when $m_\phi \gg v$,
\be
\frac{\Gamma (\phi \rightarrow \bar \chi \bar \nu h)}{\Gamma (\phi \rightarrow \bar \chi \bar \nu)} \simeq \frac{m_\phi^2}{24 \pi^2 v^2}
\ee
This three-body decay mode is strongly constrained by BBN because of the hadronic decays of the Higgs \cite{Kawasaki:2004qu, Jedamzik:2006xz}, implying that $\phi$ must decay faster than about a second if the branching fraction to three-body decays is appreciable.

Recall that in the asymmetric DM scenario, $\chi$ and $\bar \chi$ annihilate with a large enough cross-section, $\sigma_{ann}$, such that the symmetric component of DM is subdominant at the present day,
\be
\label{eq:DarkSigma}
\left< \sigma_{ann} v \right> \gg 3 \times 10^{-26} \unit{cm}^3\unit{s}^{-1} \,.
\ee
We must make sure that $\phi$ decays late enough such that these $\chi$ annihilations are decoupled (and furthermore, late enough to avoid their recoupling after the decay), so that the $\chi$ abundance is determined by the asymmetry, instead of the annihilation rate.  This decoupling temperature is defined by the relation, $s \, Y_{\Delta \chi} \left< \sigma_{ann} v \right> = H(T_{dec})$, which implies that,
\be
T_{\rm dec} =   \unit{GeV}  \frac{m_\chi}{100  \unit{GeV}} \, g_*^{-1/2}   \left(\frac{10^{-24} \unit{cm}^3/\mathrm{sec}}{\left< \sigma_{ann} v \right>} \right), 
\ee
where we've assumed that $g_* \sim g_{*S}$ at $T_{dec}$.  Consequently,
\be
\tau_{\rm dec} = 6\times 10^{-6}\unit{sec} \ g_*^{1/2} \left(\frac{100 \unit{GeV}}{m_\chi}\right)^2 \left(\frac{\left< \sigma_{ann} v \right>} {10^{-24} \unit{cm}^3/\mathrm{sec}}\right)^2\,.
\ee
Requiring that $\phi$ decays occur after this decoupling temperature leads to a nontrivial constraint on the parameters,
\be
\label{e.dupaa}
g_*^{1/2}  \, \frac{ m_\phi}{m_\chi} \, \left(\frac{\lambda}{0.1}\right)^2\left( \frac{100\unit{GeV}}{m_\chi}\right) \left(\frac{\left< \sigma_{ann} v \right>}{10^{-24} \unit{cm}^3/\mathrm{s}} \right)^2 \left(\frac{10^9\unit{GeV}}{M_{N_1}}  \right) \left( \frac{m_\nu}{0.05 \unit{eV}} \right) \,<\, 100 ,
\ee
where we've assumed that the three-body decay mode is not dominating, for this estimate.
 
If DM is dominantly asymmetric today, then DM annihilations are suppressed at the current epoch, making it difficult to observe indirect signatures of dark matter annihilations.  But when $\bar \chi$ is restored by decays of $\phi$, DM annihilation signals are also restored and proceed at a boosted rate relative to a thermal WIMP, as in Eq.~\ref{eq:DarkSigma}.  
This provides a novel mechanism for producing a large dark matter annihilation rate today.  

In this scenario, such annihilations may account for the leptonic cosmic ray excesses observed by PAMELA~\cite{Adriani:2008zr}, FERMI~\cite{Abdo:2009zk}, and HESS \cite{Aharonian:2009ah}.  The requirement is simply that DM has a TeV scale mass and that its annihilations produce leptons.  This provides an attractive alternative to the scenario where the DM annihilation rate experiences a Sommerfeld enhancement at low velocities~\cite{ArkaniHamed:2008qn}.  In models with a Sommerfeld enhancement, there can be tension between producing the correct relic density and a large enough annihilation rate in our Galaxy~\cite{Feng:2010zp, Finkbeiner:2010sm}.  But in our framework, this tension is resolved because the DM abundance follows from the asymmetry produced by leptogenesis, not from the annihilation rate at decoupling.  We note that both models with Sommerfeld enhancement, and our framework, are constrained by the Cosmic Microwave Background (CMB)~\cite{Slatyer:2009yq} (and other astrophysical constraints, see e.g.~\cite{Meade:2009iu}) because $\phi$ decays before the time of recombination.

Finally, we emphasize that decays of $\phi$ provide a mechanism for generating mixed warm/cold DM when there is a mass hierarchy: $m_\phi \gg m_\chi$.  In this regime, $\bar \chi$ is produced carrying a large kinetic energy set by $m_\phi$.  If the decay occurs late enough, such that $\bar\chi$ is kinetically decoupled, then it will not thermalize and there will not be enough time to redshift away its kinetic energy.  This opens up the possibility for $\chi$ to constitute cold dark matter while $\bar \chi$ is warm.  Parametrically, the velocity of $\bar \chi$ at the present epoch is given by~\cite{Jedamzik:2005sx},
\be
v_{\bar \chi} \sim 2\times10^{-5} \frac{\unit{km}}{\unit s} \left( \frac{m_\phi}{m_\chi} \right)  \sqrt \frac{\tau_\phi}{1\unit{s}},
\ee
where $\tau_\phi$ is the $\phi$ lifetime.  For example, if $m_\phi = 100 \unit{GeV}$,  $m_\chi = 1 \unit{MeV}$, and $\tau_\phi$ is one second, then $\bar \chi$ will have a free-streaming velocity of about 1 km/sec, large enough to impact the matter power spectrum~\cite{Jedamzik:2005sx}.  These parameters have some tension with the limit of Eq.~\ref{e.dupaa}, but the limit can be satisified at small $\lambda$ and large $M_{N_1}$.  This scenario of mixed warm/cold dark matter, where half of dark matter is warm and the other half cold, may have interesting phenomenological consequences for structure formation, which would be worthwhile to further explore.

\subsection{Asymmetric Sterile Neutrinos from Leptogenesis}\label{sec:sterile}

So far, we have assumed that the scalar in the hidden sector, $\phi$, does not obtain a VEV at low-energy.  This permits DM to be stable when $m_\phi > m_\chi$.  In this section, we relax this assumption by allowing $\phi$ to receive a nonzero VEV\@.  We will see that this simple change leads to several new phenomenological possibilities.
Because of the nonzero $v_\phi \equiv \left< \phi \right>$, DM now mixes with the left-handed neutrinos.  Therefore, it constitutes a Dirac sterile neutrino (we continue to assume that DM has a Dirac mass, which is necessary for its abundance to be set by an asymmetry). This scenario thus provides a novel mechanism to account for the correct relic abundance of sterile neutrino DM (for a nice review and references see Ref.~\cite{Kusenko:2009up}).  DM stability is no longer guaranteed, and several decay modes open up due to the mixing with neutrinos.  For the appropriate DM lifetime, this leads to observable cosmic rays at the present epoch.  Another consequence of $v_\phi\neq0$ is that DM inherits a small Majorana mass, $\mu_\chi \chi^2$, where $\mu_\chi \ll m_\chi$.  As we discuss below, this leads to oscillations at late times, allowing for a large annihilation rate at the present day, as in section~\ref{sec:PhiDecay}.  We consider the above effects in detail below.

Recall that the seesaw Lagrangian (here simplified to the one-flavor case) is given by, 
\be
\cL \supset -m_\chi \chi \chit + \frac{1}{2}M_{N_1} N_1^2 + \lambda \, N_1 \chi \left< \phi \right> + y \, N_1 L \left< h \right> + h.c. \,,
\ee
where  have included an explicit Dirac mass for $\chi$, and we emphasize that both $\phi$ and the SM higgs, $h$, receive VEVs.   After integrating out the heavy right-handed neutrino, $N_1$, we have the following mass terms,
\be
\cL \supset -m_\chi \, \chi \chit - \frac{\mu_\chi}{2}  \chi^2 - \frac{m_{\nu}}{2} \nu^2 - \mu_{\chi \nu} \, \chi \nu+ h.c. \,,
\ee
where $\mu_\chi \ll m_\chi$ constitutes a small Majorana mass for $\chi$, $m_{\nu}$ is the usual Majorana mass for left-handed the neutrino, and $\mu_{\chi \nu}$ represents a mass-mixing between $\chi$ and $\nu$.  These masses are given by,
\be \label{eq:masses}
\mu_\chi = \lambda^2  \frac{v_\phi^2}{M_{N_1}}\,, \qquad m_\nu = y^2 \frac{v_{\rm EW}^2}{M_{N_1}}\,, \qquad \mu_{\chi \nu} = \left( \frac{\lambda}{y}\frac{v_\phi}{v_{\rm EW}}\right) m_\nu\,.
\ee
The $\chi$ Majorana mass, $\mu_\chi$, leads to DM particle/antiparticle oscillations $\chi \leftrightarrow \chit$.  Similarly, the DM-neutrino mass mixing, $\mu_{\chi \nu}$, leads to DM/neutrino oscillations, $\chi \leftrightarrow \bar \nu$.  We now discuss how these oscillations can modify the cosmological history of this model.

We begin by considering only the $\chi \leftrightarrow \chit$ oscillations (we will see below that the $\chi \leftrightarrow \bar \nu$ oscillations can be neglected for the parameters of interest). It is important that the oscillations do not turn on until after DM annihilations decouple, because otherwise the $\chi$ asymmetry, resulting from leptogenesis, is erased.   As a consequence, we shall now see that the rate of oscillations are slow relative to the Hubble rate at all times.  Nonetheless, the probability to oscillate becomes sizable at late times, thereby enabling a  large annihilation rate at the present day, as in Eq.~\ref{eq:DarkSigma}.  This provides an alternative mechanism, compared to the late $\phi$ decays discussed in section~\ref{sec:PhiDecay}, for generating large cosmic ray fluxes at the present epoch~\cite{Cohen:2009fz, Cai:2009ia}.

In order to verify that DM does not oscillate too soon, we briefly review the formalism for treating particle oscillations in the expanding universe~\cite{Barbieri:1989ti, Dodelson:1993je, Abazajian:2001nj}.  Consider a generic oscillation of the type $\alpha \to \beta$.  The BE for production of $\beta$, through oscillations, is given by,
\be
\frac{dY_\beta}{d z} = {z \over 2} \left< P_{\alpha \to \beta}\left( t \right)\right>  \frac{\Gamma_{\alpha}}{H_1} \left(Y_\alpha - Y_\beta \right) \,,
\ee
where $P_{\alpha \to \beta}\left( t \right)$ is the probability that $\alpha$ oscillates into $\beta$ after time $t$, $\Gamma_{\alpha}$ is the total interaction rate of $\alpha$, $z = m/T$ is defined in terms of an arbitrary mass scale $m$, and $H_1\equiv H(T=m)$.  The oscillation probability, $P$, is averaged over the interaction time, $\left< P \right> = \Gamma_{\alpha} \int_0^\infty dt \, e^{-  \Gamma_{\alpha} t} \, P$.  We see that the BE is driven by $P \times \Gamma$.  Therefore, oscillations are in equilibrium whenever $P \Gamma   \gg H$ and are frozen out whenever $P \Gamma \ll H$.  The general expressions for $P$ and $\left< P \right>$ are:
\be \label{eq:Pdef}
P_{\alpha \to \beta}\left( t \right) = \sin^2 \left(2 \, \theta_{\alpha \beta}\right) \, \sin^2 \left(\frac{ \Delta E_{\alpha \beta}}{2} \, t \right) \,, \qquad 
 \left< P_{\alpha \to \beta}\left( t \right)\right>  = \frac{\sin^2 \left(2 \, \theta_{\alpha \beta}\right)}{2}\frac{\Delta E_{\alpha \beta}^2}{\Delta E_{\alpha \beta}^2+\Gamma_\alpha^2}\,,
\ee
where $\theta_{\alpha \beta}$ is the mixing angle and $\Delta E_{\alpha \beta} = t_{osc}^{-1}$ the energy difference between the states $\alpha$ and $\beta$, or equivalently the inverse oscillation time.  Note that for simplicity, we neglect the effects of finite temperature and density.  These corrections are considered, for example, in many studies of sterile neutrino production through neutrino/sterile neutrino oscillations~\cite{Barbieri:1989ti, Dodelson:1993je, Abazajian:2001nj}.

We now apply the above formalism to understand $\chi \leftrightarrow \chit$ oscillations.  
The mixing angle and the energy splitting in this case are given by 
\be \label{eq:OscThetaE}
\theta_{\chi \chit} \simeq {\pi \over 4} \,,
\qquad
 \Delta E_{\chi \chit} \simeq  \left \{ \ba{cc}  
 \mu_\chi m_\chi /T & T > m_\chi
 \\
 \mu_\chi & T \le m_\chi  
 \ea \right .
\ee
In order to insure that these oscillations are slow before DM annihilations decouple, it is sufficient to require that $\Gamma P < H$ at the temperature of decoupling,
\be \label{eq:OscLimit0}
 \left< P_{\chi \rightarrow \chit}\left( t \right)\right> \Gamma_\chi(T_{\mathrm {dec}}) \, \lesssim \, H(T_{\mathrm {dec}}). 
\ee
Since the asymmetric component of $\chi$ cannot be neglected, $T_{\rm dec}$ must be calculated by comparing the total number density times the annihilation rate to Hubble, $\Gamma_{\rm ann} = [n_{\rm asym}(T_{\rm dec}) + n_{\rm sym}(T_{\rm dec})]\langle \sigma_{\rm ann}v\rangle = H(T_{\rm dec})$.  For our case, where the asymmetric component dominates at low temperatures one finds, $m_\chi / T_{\rm dec}\gg 20$ implying that even at temperatures below that at which the symmetric component decouples, $m_\chi/T_{\rm dec}^{\rm sym}\sim 20$, oscillations may recouple the annihilation process, thereby significantly altering the DM abundance.  
Using Eqs.~\eqref{eq:Pdef} and \eqref{eq:OscThetaE}, Eq.~\eqref{eq:OscLimit0} reads at $T=T_{\rm dec}$,
\be
\mu_\chi \lesssim \sqrt{\Gamma_\chi \Gamma_{\rm ann}({T=T_{\rm dec}})} \,.
\ee
This condition can be converted into a limit on the size of $\lambda \, v_\phi$, which enters the oscillation probability through $\mu_\chi$. 
Using Eqs.~\ref{eq:masses} one finds the limit,
\be \label{eq:OscLimit1}
\frac{\lambda \, v_\phi}{m_\chi} \, \lesssim \, 3\times10^{-7} \left({M_{N_1} \over 10^{10}\unit{GeV} }\right)^{1/2} \left(10^{-24}\unit{cm}^3/\mathrm{s} \over \left< \sigma_{ann} v \right>\right)  \left( \left< \sigma_{tot} v \right> \over \left< \sigma_{ann} v \right> \right)^{1/4} \left(\frac{g^{3/4}_*}{g_{*S}}\right)\,,
\ee
where $\sigma_{ann}$ is the $\chi+ \bar \chi$ annihilation cross-section, $\sigma_{tot}$ is the overall $\chi$ interaction cross-section (which determines the total DM interaction rate $\Gamma_\chi$), and $g_*$ and $g_{*S}$ are evaluated at the temperature that DM annihilations decouple.
We see that the requirement that DM particle/antiparticle oscillations are slow, before DM decouples, presents a rather stringent limit on the quantity $\lambda \, v_\phi$ relative to the DM mass, $m_\chi$.

We note that if Eqs.~\ref{eq:OscLimit0} and~\ref{eq:OscLimit1}, are satisfied, then $P \Gamma \ll H$, also at temperatures below DM decoupling, due to the drop in $\Gamma$ relative to $H$ as the Universe cools.  
This means that, given the condition  Eq.~\ref{eq:OscLimit0} is satisfied, oscillations always remain slow in the expanding universe,   and never produce a significant yield of $Y_{\chit}$.  
Still, there is a large annihilation rate today, as long as $t_{osc} = \mu_\chi^{-1}$ is much shorter than the age of the universe.  In this regime, whenever two DM particles collide at the present day, there is an $\cO(1)$ probability that one has oscillated.  We also comment that if $\mu_\chi$ is chosen so that $t_{osc}$ is longer than the timescale for recombination, then the constraints from the CMB~\cite{Slatyer:2009yq} are alleviated. 

Now we consider oscillations between DM and neutrinos, $\chi \leftrightarrow \bar \nu$.  Oscillations of this type are highly suppressed by the small mixing angle between DM and neutrinos,
\be \label{eq:NuOscThetaE}
\theta_{\chi \nu} \simeq {\mu_{\chi \nu} \over m_\chi}\,,
\qquad
\Delta E_{\chi \nu} \simeq  \left \{ \ba{cc}  
 m_\chi^2 /T & T > m_\chi
 \\
 m_\chi & T \le m_\chi  
 \ea \right .
\ee
In particular, combining the constraint of Eq.~\ref{eq:OscLimit1} with Eqs.~\ref{eq:masses}, \ref{eq:Pdef}, and \ref{eq:NuOscThetaE}, we learn that,
\be
\left< P_{\chi \rightarrow \bar \nu}(t) \right> \, \simeq \, 2 \left(\frac{\mu_{\chi \nu}}{m_\chi}\right)^2 \, \lesssim \, 10^{-33} \left( m_\nu \over 0.05\unit{eV}\right) \left( \frac{10^{10}\unit{GeV}}{M_{N_1}}\right)
\ee
This oscillation probability is small enough that $P \Gamma \ll H$ for all temperatures.  This shows, {\it a posteriori}, that we were justified to neglect DM/neutrino oscillations in the above discussion.

\begin{figure}[t]
\begin{center}
\includegraphics[width=3.2in]{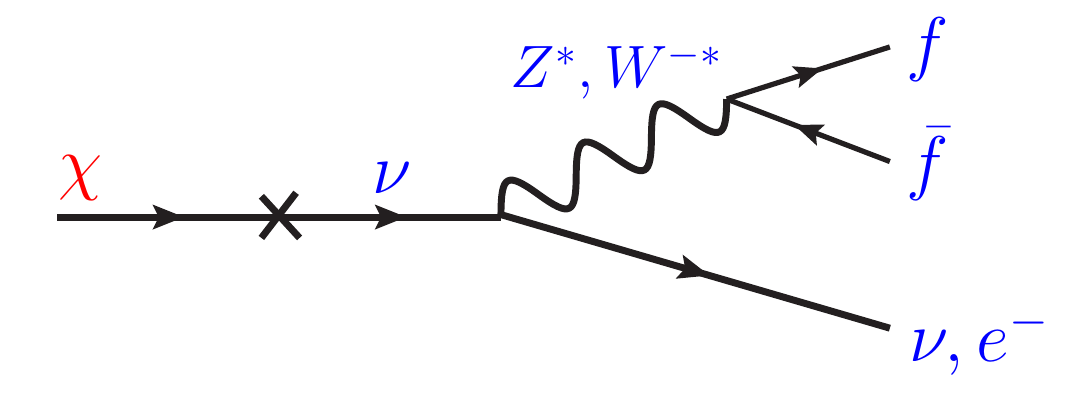}
\caption{\small\it Dark matter decay modes in the case where $\langle\phi\rangle\neq 0$.  These modes occur due to  mixing of DM with SM neutrinos and place stringent constraints on the mixing angle for $m_\chi\gtrsim 0.1\unit{GeV}$\@.  For lighter DM, oscillations into $\chit$ are more constraining.   }
\label{fig:chidecay}
\end{center}
\end{figure}

There is, however, one very important effect of DM/neutrino mixing, $\theta_{\chi \nu}$.  This mixing opens up several decay modes for DM\@.  For example, suppose that $\chi$ is lighter than the electroweak gauge bosons.  Then as shown in Fig.~\ref{fig:chidecay}, DM can three-body decay through off-shell $Z^*/W^*$, $\chi \rightarrow \nu (Z^*\rightarrow f \bar f)$ and $\chi \rightarrow e^- (W^{+*}\rightarrow f \bar f')$, where $f / f'$ are SM fermions.
The rate for these decays is given, parametrically, by the expression,
\be \label{eq:3bod}
\Gamma_{\chi \rightarrow \nu f \bar f} \simeq \frac{\alpha_2^2 \, \theta_{\chi \nu}^2}{4 \pi} \, \left(\frac{m_\chi}{m_Z}\right)^4 \, m_\chi
\ee
Of course, these decays are constrained to not proceed too quickly.  In order to be consistent with cosmic ray and diffuse gamma observations, $\tau_\chi \gtrsim 10^{26}$~s~\cite{Meade:2009iu, Papucci:2009gd}.  This limit can also be expressed as a constraint on the size of $\lambda \, v_\phi$ relative to $m_\chi$, 
\be
\frac{\lambda \, v_\phi}{m_\chi} \, \lesssim \, 10^{-9} \left(\mathrm{GeV} \over m_\chi \right)^{5/2}  \left({M_{N_1} \over 10^{10}\unit{GeV} }\right)^{1/2} \left(\frac{0.05\unit{eV}}{m_\nu} \right)^{1/2} \left( \frac{10^{26}\unit{s}}{\tau_\chi} \right)^{1/2}
\ee
We see that the DM lifetime is a stronger constraint than DM particle/antiparticle oscillations, when $m_\chi \gtrsim 0.1$~GeV, while oscillations present the dominant constraint on light DM\@.  For DM above the GeV scale, the cosmic rays are observable, by ongoing and future observations, when $\lambda \, v_\phi$ saturates the above limit.  We note that sterile neutrino DM of this type can also decay, at one loop, to a photon line, $\chi \rightarrow \nu \gamma$~\cite{Kusenko:2009up}.  However, this decay is suppressed, relative to the above three-body decays, by a factor of $\sim \alpha/4 \pi$.  We also note that in the regime where $m_\chi > m_Z$, DM will dominantly decay two-body to longitudinal electroweak gauge bosons and the Higgs, $\chi \rightarrow Z \nu$, $\chi \rightarrow W^+ e^-$, and $\chi \rightarrow h \nu$.  In this regime, the decays proceed faster than above, and the constraint on $\lambda \, v_\phi$ is significantly stronger.

\section{Cosmology and Light Dark Matter}\label{sec:cosmo}

As we have seen, asymmetric DM populated by leptogenesis can accommodate a wide range of DM masses,  between about keV and $10$~TeV\@.  
Within this broad framework, cosmology, astrophysics, and colliders can all constrain the properties of DM, and the hidden sector within which it resides.  In particular, the hidden sector must contain additional degrees of freedom, lighter than DM, which the symmetric DM component can annihilate into.
While the constraints on weak-scale DM are well-known (for a review see~\cite{Jungman:1995df}), light (keV to $10$ GeV) DM raises several interesting issues, which are the subject of this section.

For concreteness, we specialize to models with a hidden $U(1)_d$ gauge symmetry~\cite{Cohen:2010kn}, where the symmetric DM component annihilates predominantly into a pair of hidden sector photons, $\gd$.
The hidden sector couples to the visible sector through the vector portal, 
\begin{equation}
{\cal L} \supset \frac{\epsilon}{2} F^\prime_{\mu\nu}F^{\mu\nu}\,,
\end{equation}
where $F^\prime_{\mu\nu} (F^{\mu\nu})$ is the $\gd$ (electromagnetic) field strength.  
The above coupling allows $\gd$ to decay into kinematically available SM fields with non-zero electric charge.  
The width and lifetime of $\gd$ is determined  by its mass $m_{\gd}$ and by the mixing parameter $\epsilon$.

We will find it helpful to consider two cases separately, depending on whether $m_\chi$ is heavier than, or lighter than, the MeV scale.
First, we consider $m_\chi \gtrsim$~MeV\@.  In this regime, we are also free to assume that $m_{\gamma_d} \gtrsim$~MeV, and in particular that no sub-MeV hidden states exist.
This means that there are no new relativistic DOF present during BBN, and the only constraint from BBN is that the the lightest state in the hidden sector decays to the SM fast enough to  avoid late dissociation processes~\cite{Kawasaki:2004qu,Jedamzik:2006xz}.  This requirement is not hard to fulfill.
For instance, if the hidden photon is the lightest hidden state, it decays to electron pairs before $T \sim $ MeV for~\cite{Ruderman:2009tj}
\begin{equation}
\label{eq:BBN1}
\epsilon > 3 \times 10^{-11} \left(\frac{{\rm GeV}}{m_{\gamma_d}}\right)^{1/2}\,.
\end{equation}
Further constraints on $\gamma_d$ are summarized in~\cite{Jaeckel:2010ni}.  
For a hidden photon heavier than $\sim$ GeV, the constraints come from B-factories and muon anomalous magnetic moment and imply $\epsilon \lesssim 10^{-3}$.   
For  $m_{\gd}$  between MeV and GeV, beam dump experiments (typically assuming a 2-body decay of the lightest state) require $\epsilon \lesssim 10^{-7}$.  
Finally, for $m_{\gd}$ between  $\sim1-50$ MeV, constraints from supernovae cooling imply $\epsilon \lesssim 4\times 10^{-9}$. 

\begin{figure}[t]
\begin{center}
\includegraphics[scale=0.65]{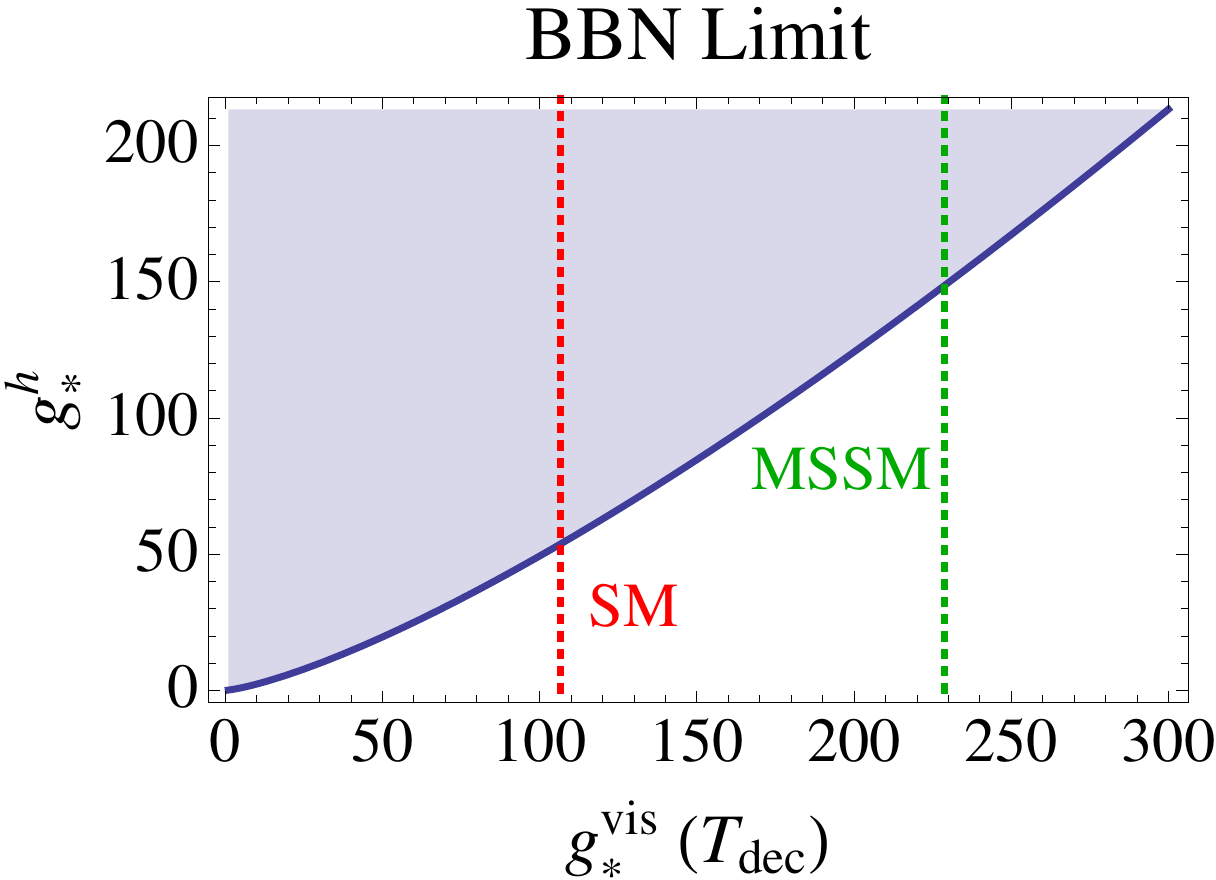}
\caption{\it\small The BBN limit on the effective number of degrees of freedom in the hidden sector, $g_*^h$, as a function of the number of degrees of freedom in the visible sector, $g_*^{vis}$, at the temperature the two sectors decouple, $T_{dec}$.  For reference, the vertical dashed lines indicate the sizes of the full SM and MSSM\@.  Here we have assumed that the entire hidden sector is lighter than the BBN scale $\sim$ MeV\@.}
\label{fig:mooncake}
\end{center}
\end{figure}

Now we consider very light DM, $m_\chi \in [$keV , MeV$]$\@. In this regime, there are two important changes to the above discussion: (1) there are additional light DOF present during BBN, and (2) for $m_{\gamma_d} < 2 m_e$, the hidden photon becomes cosmologically long-lived because there remain no lighter states with electric charge.  Both of these facts are potentially hazardous, and we now discuss how to evade the danger.

The model can be ruled out by BBN, because DM plus the hidden photon exceed the number of relativistic DOF allowed during BBN at the SM temperature.  This constraint can be avoided if the hidden sector is cooler than the SM when $T \sim$~MeV\@.  This is only possible if kinetic equilibrium is not maintained between the two sectors, which requires that the following type of reaction be inefficient: 
$\gd \, e^\pm \leftrightarrow \gamma \, e^\pm$.  This reaction is decoupled when $T \gtrsim (\epsilon^2 \alpha_{\rm EM}^2 / \pi^2 g_*^{1/2}) M_{\rm Pl}$.
Therefore, the two sectors are not in kinetic equilibrium at $T_{BBN} \sim 1$ MeV if  
\begin{equation}
\label{eq:epsilondec}
\epsilon <  \epsilon_{\rm BBN}  \simeq  7\times10^{-9} \,. 
\end{equation}
Suppose that this condition is satisfied, and that the two sectors begin at the same temperature (for example at the leptogenesis scale) and decouple at a lower temperature, $M_{N_1} > T_{dec} > T_{\rm BBN}$.  If more DOF freezeout in the SM than the hidden sector, before BBN, then by the separate conservation of entropy in the two sectors, $T_h < T$.
The limit from BBN becomes, at $95\% $ CL~\cite{Feng:2008mu},
\begin{equation} 
g_*^h(T^h_{\rm BBN})\left(\frac{T^h_{\rm BBN}}{T^{}_{\rm BBN}}\right)^4 < 2.52 \,.
\end{equation}
Here $T^h_{\rm BBN}$ is the temperature in the hidden sector at BBN, while $g_*^h(T)$ is the effective number of relativistic degrees of freedom in the hidden sector, as defined in~\cite{Feng:2008mu}.  
In Fig.~\ref{fig:mooncake} we use this relation to show the constraint on the size of the hidden sector, $g_*^h$,  as a function of the number of DOF in the SM when the two sectors decouple, $g_*^{vis} (T_{dec})$, assuming that all hidden particles are lighter than the temperature of BBN\@.  We see that rather large hidden sectors can be accommodated, $g_*^h \sim 50$ if the two sectors decouple above the electroweak scale.

The bottom line of the above discussion is that the two sectors cannot be in thermal equilibrium at the time of BBN, which enforces the constraint $\epsilon < \epsilon_{BBN}$.
We note that even stronger constraints arise from  bounds on the lifetime of the sun and horizontal branch stars that require  $\epsilon\lesssim 10^{-13}$~\cite{Jaeckel:2010ni}.

Finally, we discuss how to avoid overclosing the Universe, since the hidden photon is cosmological long-lived below the MeV scale.
One possibility is to drop the assumption that the hidden sector has only one mass scale. 
Then we can make the lightest state in the hidden sector sufficiently light, such that it  does not overclose the universe even if it is cosmologically stable\footnote{The lightest state may reside in a distinct sector thus explaining the hierarchy of scales.}.  An estimate for the upper bound on the mass of the lightest state, $\hp$,  can found by assuming it follows a thermal distribution.   At temperatures above its mass, $T^h > m_{\hp}$, the number density is related to the photon number density through,
\begin{equation}
n_{\hp}(T^h) = C_{\hp}\frac{g}{2}\left(\frac{T^h}{T}\right)^3 n_\gamma(T)\,,
\end{equation}
with $C_{\hp}=3/4$  ($C_{\hp}=1$) for a fermion (boson) $\hp$.  Below $m_{\hp}$, the energy density scales as $a^{-3}$ and is therefore proportional to $n_\gamma(T_0)$.   Comparing to the measured DM energy density, one finds,
\begin{equation}
\label{eq:mhpbound}
m_{\hp} \leq 3 \textrm{ eV} \left(\frac{2}{g}\right)C_{\hp}^{-1} \xi_{\hp}^{-3}\,,
\end{equation}
where $\xi_{\hp} = T^h / T$ at the time when $\hp$ becomes non-relativistic.  
This result is in agreement with the more precise calculation of \cite{Das:2010ts}. 

Another intriguing possibility exists.  
Above we assumed that the hidden sector was in thermal equilibrium with the SM at the time of leptogenesis.
However if the hidden sector couples only weakly to the right-handed neutrino, the two sectors may have never been in equilibrium with each other.  
In that  case the hidden sector can easily be much colder than the SM,  so as to allow for a heavier $m_{\hp}$, cf. Eq.~\eqref{eq:mhpbound}.  
Furthermore, it is interesting to note that if the DM annihilation rate is slow (or even vanishing), the symmetric component can dominate over the asymmetric one. Then the symmetric component would be responsible for  the observed relic abundance, while at the same time it would be related to generation of the baryon asymmetry. 
This possibility is however outside the ADM paradigm, therefore we postpone its study to future work~\cite{futurework}.

As a final remark, we note that the above discussion assumed a hidden gauge group coupled to the visible sector through gauge kinetic mixing.  Other portals, such as the Higgs portal may be considered.  The Higgs portal can have very different (and potentially weaker) astrophysical constraints than the vector portal.   This is because $\gamma_d$ couples to electric charge and therefore equally to electrons and protons, whereas a hidden sector scalar, that couples through the Higgs portal, will couple more strongly to protons than electrons.

\section{Outlook}\label{sec:conclusions}

In this paper we discussed the ADM scenario in the context of 2-sector leptogenesis. ÊThe asymmetries in the SM and DM sectors depend on several factors, such as the branching fractions and decay asymmetries of the sterile neutrinos, and the strength of washout effects in each sector.
Consequently, the ratio of dark matter number density to the baryon number density is very sensitive to the model parameters, thereby accommodating 
a wide range of Êdark matter masses from keV to 10 TeV\@. 
 ÊOur findings suggest that the spectrum of predictions of ADM, especially of those concerning the dark matter mass, Êis much wider than previously thought. 
This is very important for planning direct and indirect experimental searches that target the ADM scenario. 

Here, we have chosen to focus on the concrete scenario of thermal leptogenesis with hierarchical neutrino masses. ÊThere are several variations and open questions that remain unexplored:
\begin{itemize}
\item Other leptogenesis scenarios, such as Êsoft leptogenesis, resonant leptogenesis or ÊDirac leptogenesis, may be accommodated within the 2-sector leptogenesis framework. Ê It would be interesting to study the above variations as they are Êexpected to admit different dynamics and in some cases produce very different phenomenology. 
\item It would be interesting to extend our treatment of the BEs to include finite temperature corrections and the full flavor structure of the theory. ÊThese corrections are well-studied for traditional leptogenesis and may have interesting consequences for 2-sector leptogenesis as well.
\item In one of the variations studied here, the DM mixes with the active neutrinos, providing a novel realization of Êsterile neutrino DM\@. ÊIt is worthwhile to investigate this scenario further. ÊIn particular, it would be interesting to understand how the constraints derived from late time oscillations or decays can be ameliorated in other leptogenesis scenarios. Ê
\item It is possible to imagine a similar mechanism for populating ADM in a single-sector leptogenesis model~\cite{cliff}. ÊClearly the phenomenology of such a scenario would be distinct.
\item Another deformation studied here predicts Êdark matter to be a mixture of cold and warm components with Êroughly equal numbers. 
This Êmay have interesting phenomenological consequences for structure formation, which would be worthwhile to explore.
\item It may also be interesting to consider variants of this scenario where the symmetric DM component dominates the relic density. 
In particular, the symmetric component originally produced by the decays of the heavy sterile neutrinos may not annihilate away.
In such a case, the asymmetric lepton density is suppressed (due to the usual bounds on the decay asymmetry into the SM sector) with respect to the DM number density, thereby predicting light DM\@. Ê
We postpone a study of the details of such scenarios to future work~\cite{futurework}.
\end{itemize}

\section*{Acknowledgments} 
We thank Ami Katz for collaboration in the early stages of this project.
We also thank Cliff Cheung, Tim Cohen, Yanou Cui, Rouven Essig, Yuval Grossman, Marc Kamionkowski, Yasunori Nomura, Michele Papucci, Aaron Pierce, and Tracy Slatyer for useful conversations.  The work of T.~V. was supported in part by the Director, Office of Science, Office of High Energy and  Nuclear Physics, of the US Department of Energy under Contract DE-AC02-05CH11231.

\appendix

\section{Appendix: Boltzmann Equations for the 2-Component Toy Model }\label{sec:appendix}

For completeness, we discuss the BEs in the  2-component toy model, taking into account the 2-to-2 washout and transfer terms. 
The equations for the asymmetries $Y_{\Delta a} =   Y_{a} -  Y_{\bar a} $ of the components $l$ and $\chi$ read 
\begin{eqnarray}
\label{e.2ct_fullbe}
{d Y_{\Delta a}  \over dz} &=&  - \GoH 
\Bigg [\epsilon_a  {z K_1(z) \over K_2(z)}  (Y_{N_1}^{eq}- Y_{N_1})  
+ 2  {\br}_a^2 I_W(z)  Y_{\Delta a} 
 \nonumber \\ &&
+  {\br}_a   {\br}_b  I_{T_+}(z) (Y_{\Delta a}+Y_{\Delta b} ) 
+  {\br}_a   {\br}_b  I_{T_-}(z) (Y_{\Delta a} -Y_{\Delta b} )
 \Bigg], 
\end{eqnarray}
where $a = l,\chi$, $b = \chi,l$, and  $H_1 = H(T=M_{N_1})$. 
The source term proportional  to the  decay asymmetry of $N_1$ is the same as in the narrow-width approximation we discussed earlier. 
The second term is due to washout processes $a a \leftrightarrow \bar a \bar a$ within one component, and the terms in the second line are due to transfer processes  $a a \leftrightarrow \bar b \bar b$ and  $a a \leftrightarrow b b$ between the 2 components.  
Explicitly, the washout and transfer terms take the form 
\be
I_i(z)  =    {\hat \Gamma \over \pi}  \int_0^\infty dt t^2 K_1(t)  f_i(t^2/z^2) ,   
\ee
where $\hat \Gamma =  \Gamma_{N_1}/M_{N_1}$ and  
\be
\label{e.amplitudeFunctions}
f_{W}(s) &=& {s/2 \over (s-1)^2 +  \hat \Gamma^2} + {s - \log(s+1) \over s}    - {2 (s - 1) \over  (s-1)^2 +  \hat \Gamma^2}  {(s+1) \log(s+1) - s \over s} 
 \nonumber \\  &&
+ {s/2 \over s +1} +  {\log(s+1)\over s+ 2 },
 \nonumber \\  
f_{T_+}(s) &=& {s/2 \over (s-1)^2 + \hat \Gamma^2} + {s \over s +1}  + {s - \log(s+1) \over s},
\nonumber \\  
f_{T_-}(s)  &=& {s^2/2 \over (s-1)^2 +   \hat  \Gamma^2}  + {(s+1) \log(s+1) -s \over s + 1 } + {(s + 2) \log(s+1) - 2 s \over s}.
\ee 
In the limit $\hat \Gamma \to 0$ the pole terms in $f_i$ dominates the integral at small and moderate $z$, and we find $I_{i} \approx  z^3 K_1(z)/4$. 
The BEs then reduce to 
\be 
\label{e.2ct_strumiabe}
{d Y_{\Delta a}  \over dz} =  - \GoH 
\left [\epsilon_a  {z K_1(z) \over K_2(z)}  (Y_{N_1}^{eq}- Y_{N_1})  +  {\br}_a  {z^3 K_1(z) \over 2} Y_{\Delta a}  \right],
\ee
where the second term describes  washout out due to $2 \to 1$ inverse decays. 
We see that in the limit  $\hat \Gamma \to 0$ the two components decouple and evolve independently. 
Their final asymmetries are determined by the decay asymmetries $\epsilon_a$ and the strength of washout effects  in each sector  set by $ {\br}_a \goh$.  In addition, the asymptotic value for the asymmetry depends on the initial conditions for the sterile neutrino yield $Y_{N_1}$, when washout is weak.  See the discussion in Sec.~\ref{sec:washout}.

As the pole contributions to the integrals decay exponentially with $z$,  
the asymptotic behavior of the BEs is determined by ${\cal O}(\hat \Gamma)$ off-pole contributions who are only power suppressed at large $z$. 
In other words, at late times $2 \to 1$ inverse decays become subdominant with respect to 2-to-2 off-pole processes. 
The crossover typically happens for moderate $z$, around $z_c \sim 25$, where $z_c$ only logarithmically depends on $\hat \Gamma$.  
For $z > z_c$ we can approximate
 \be
I_D(z) \simeq  3 {W \over  z^2} \HoG , 
\qquad 
I_{T_+}(z) \simeq   {W \over  z^2} \HoG , 
\qquad 
I_{T_-}(z) \simeq  14  {W \over  z^4} \HoG,    
\ee   
where $W = (32 \hat \Gamma/\pi)(\goh)$ is the order parameter that sets the strength of the late time washout and transfer effects.  
For $z  > z_c$  the BEs reduce to  
\be
{d  \over dz} \left [ \begin{array}{c}  Y_{\Delta l} \\ Y_{\Delta \chi} \end{array}  \right ] =  
- {W \over z^2}  M  \cdot \left [ \begin{array}{c}  Y_{\Delta l} \\ Y_{\Delta \chi} \end{array}  \right ] 
\quad 
M  = \left ( \begin{array}{cc} 6 {\br}_l^2 +  {\br}_l { \br}_\chi  & {\br}_l   {\br}_\chi \\ 
 {\br}_l   {\br}_\chi  & 6 {\br}_\chi^2 +  {\br}_l { \br}_\chi   
\end{array} \right ) .   
\ee 
Thus, the asymptotic asymmetry can we written as   
\be
 \left [ \begin{array}{c}  Y_{\Delta l}^\infty \\ Y_{\Delta \chi}^\infty \end{array}  \right ]  = \left [\exp \left ( - {W \over z_c} M \right)  \right ] \cdot 
 \left [ \begin{array}{c}  Y_{\Delta l} (z_c) \\ Y_{\Delta \chi}(z_c) \end{array}  \right ] 
\ee
in terms of the ``boundary conditions" at the crossover scale $z_c$.   
It is illuminating to study the limit where the branching fraction into one sector, say ${\rm BR}_l$,  is very small, 
in which case the exponent simplifies to 
\be
 \left [\exp \left ( - {W \over z_c} M \right)  \right ]  \simeq 
\left ( \ba{cc} e^{-  {\br}_l W /z_c}     & - {{\br}_l  \over 6}  e^{-  {\br}_l W/z_c} \left ( 1 -  e^{- 6 W /z_c} \right) \\ 
- {{\br}_l  \over 6}  e^{-  {\br}_l W/z_c} \left ( 1 -  e^{- 6W/z_c} \right) & {{\br}_l ^2 \over 36} +  e^{- 6 W/z_c}  
\ea \right ).   
\ee
For $W/z_c  \ll 1$ the diagonal elements are approximately $1$, and then late washout does not affect the asymmetries.
In that case the washout effects are very well described by the $2 \to 1$ processes encoded in Eq.~\ref{e.2ct_strumiabe}.  
The magnitude of the off-diagonal terms describing transfer is ${\br}_l W /z_c \ll 1$;  nevertheless, transfer could still be relevant in this limit if    $Y_{\Delta l} (z_c) \gg Y_{\Delta \chi}(z_c)$, or the other way around. 
  Late washout and transfer become important when $W/z_c \gtrsim 0.1$, that is when $\hat \Gamma \goh \gtrsim 0.1$. 
For ${\br}_l  \ll 1$ we typically expect $Y_{\Delta l} (z_c) \gg Y_{\Delta \chi}(z_c)$  (unless $\epsilon_l \ll \epsilon_\chi$) because of the $2 \to 1$ washout at moderate $z$ that is larger in the $\chi$ sector.  
If this is the case, one effect is the suppression of the asymmetries by the factor $e^{-  {\br}_l W/z_c}$, which may or may not be small 
depending on  ${\br}_l \hat \Gamma \goh$.  
Another effect is  that the asymmetries become correlated, 
\be
Y_{\Delta l}^\infty/Y_{\Delta \chi}^\infty \sim - 1/{\br}_l, \qquad \frac{\hat\Gamma\, \Gamma_{N_1}}{H_1} \gtrsim 0.1 \ .  
\ee 
Note that the asymmetries end up with opposite signs: particle domination in one sector implies anti-particle domination in the other.  
Amusingly, the larger magnitude of the asymmetry survives in the sector with a smaller branching fraction for $N_1$ decays. 
Contrary to naive expectations, transfers do {\em not} lead to $Y_{\Delta l}^\infty/Y_{\Delta \chi}^\infty \sim 1$, unless ${\br}_l \sim {\br}_\chi$. 
Thus,  $Y_{\Delta l}^\infty/Y_{\Delta \chi}^\infty \sim 1$, often advertised as the prediction of ADM,  is {\em not} a generic prediction of our 2-sector leptogenesis scenario. 

\newpage

\end{document}